\documentclass[journal]{IEEEtran}
\usepackage[caption=false,font=normalsize,labelfont=sf,textfont=sf]{subfig}
\pdfoutput=1
\usepackage{cite}
\usepackage{amsmath}		
\usepackage{amsfonts}		
\usepackage{amssymb}        
\usepackage{ulem}
\ifCLASSINFOpdf
\usepackage{graphicx}
\usepackage{txfonts}
\else
\usepackage{graphicx}
\fi
\usepackage{float}
\usepackage{multicol}
\usepackage{epstopdf}
\usepackage{breqn}
\usepackage{multirow}
\usepackage{caption}
\usepackage{soul}
\usepackage{algorithm}
\usepackage{algpseudocode}
\usepackage{mathtools}
\usepackage{xcolor}

\usepackage{bbm}
\usepackage{dblfloatfix}
\DeclareMathOperator*{\argmax}{arg\,max}
\DeclareMathOperator*{\argmin}{arg\,min}

\usepackage{hyperref}
\hypersetup{
    colorlinks=true,
    linkcolor=blue,
    filecolor=magenta,      
    urlcolor=blue,
}

\pdfoutput=1

\begin{document}

\definecolor{amber(sae/ece)}{rgb}{1.0, 0.49, 0.0}
\definecolor{aqua}{rgb}{0.0, 1.0, 1.0}

\title{Multi-armed Bandit Algorithms on System-on-Chip: Go Frequentist or Bayesian?}


\author{S. V. Sai Santosh, and Sumit J. Darak
					\thanks{This work is supported by the funding received from core research grant (CRG) awarded to Dr. Sumit J. Darak from DST-SERB, GoI.}
			\thanks{S. V. Sai Santosh, and Sumit J. Darak are with Electronics and Communications Department, 
				IIIT-Delhi, India-110020 (e-mail: \{siripurapu17197,sumit\}@iiitd.ac.in)}
}
	\maketitle
\begin{abstract}
Multi-armed Bandit (MAB) algorithms identify the best arm among multiple arms via exploration-exploitation trade-off without prior knowledge of arm statistics. Their usefulness in wireless radio, IoT, and robotics demand deployment on edge devices, and hence, a mapping on system-on-chip (SoC) is desired. Theoretically, the Bayesian approach-based Thompson Sampling (TS) algorithm offers better performance than the frequentist approach-based Upper Confidence Bound (UCB) algorithm. However, TS is not synthesizable due to \textit{Beta} function. We address this problem by approximating it via a pseudo-random number generator-based approach and efficiently realize the TS algorithm on Zynq SoC. In practice, the type of arms distribution (e.g., Bernoulli, Gaussian, etc.) is unknown and hence, a single algorithm may not be optimal. We propose a reconfigurable and intelligent MAB (RI-MAB) framework. Here, intelligence enables the identification of appropriate MAB algorithms for a given environment, and reconfigurability allows on-the-fly switching between algorithms on the SoC. This eliminates the need for parallel implementation of algorithms resulting in huge savings in resources and power consumption. We analyze the functional correctness, area, power, and execution time of the proposed and existing architectures for various arm distributions, word-length, and hardware-software co-design approaches. We demonstrate the superiority of the RI-MAB over TS and UCB only architectures.

\end{abstract}
\begin{IEEEkeywords}
 Intelligent architecture, Zynq SoC, Thompson sampling, multi-Armed Bandit, partial reconfiguration
 \end{IEEEkeywords}

\section{Introduction}

Multi-armed bandit (MAB) algorithms are designed to identify the best arm among multiple arms without prior knowledge of type of arm distribution and their statistics \cite{Book12_RegretAnalysis_Bubeck,mab2,mab3,mab1,TAC1987_AsyptoticallyEfficient_AnantharamWalrand}. They achieve this by optimizing an exploration-exploitation trade-off over a finite horizon (i.e. time slots) \cite{mab1,TAC1987_AsyptoticallyEfficient_AnantharamWalrand}. Here, exploration refers to selection of all arms sufficient number of times to accurately learn their statistics and exploitation refers to selection of the best arm as often as possible. Their applications include online advertisement selection to increase the number of clicks \cite{mab4,mab5}, clinical trails to identify best drugs \cite{mab4,mab5}, news personalization \cite{mab4,mab5}, decision making in financial markets \cite{mab4,mab5}, and resource selection in wireless networks \cite{JSAC11_DistributedLearning_Anadakumar,Infocom2019_DistributedLearning_TibrewalPatchalaHanawal,JSAC2019_DistributedLearning_BistrizLeshem}, internet of things (IoT) \cite{JSAC2019_MultiPlayerBanditsStableMatching_DarakHanawal,GoT, TSP14_DistributedStochastic_GaiKrishnamachari, NeurIPS2019_SICMMAB_BoursierPerchet} and robotics \cite{robotics1,robotics2}.

The performance metric for the MAB algorithm is the regret which is proportional to the number of selection of the sub-optimal arms and it should be as low as possible \cite{Book12_RegretAnalysis_Bubeck,mab2,mab3,mab1,TAC1987_AsyptoticallyEfficient_AnantharamWalrand}. An optimal MAB algorithm guarantees a logarithmic regret which is the best one can achieve. The upper confidence bound (UCB) algorithm \cite{mab1}, Kullback Leibler UCB (KL-UCB) \cite{mab2}, and Thompson Sampling (TS) \cite{mab3} are the popular optimal MAB algorithms. The KL-UCB algorithm is computationally complex due to underlining optimization routine and hence, TS and UCB algorithms are preferred in \cite{mabarch1}.

In wireless radio, IoT, and robotics applications, MAB algorithms are used for decision-making tasks in the media-access control (MAC) layer. With tight integration of MAC and physical (PHY) layer, there is an option to accelerate the MAB and other MAC algorithms on hardware such as ASIC or FPGA instead of sequential processors based software implementation. The hardware implementation exploits the parallel architecture thereby offering lower execution time i.e. latency. Such applications also demand deployment on edge devices and hence, mapping of the MAB algorithms on system-on-chip (SoC) is desired. Recently, we discussed the implementation of UCB and KL-UCB algorithms on the heterogeneous Zynq SoC from Xilinx consisting of the dual-core ARM processor and 7-series FPGA \cite{mabarch1}. In this work, we focus on the TS algorithm which has not been realized on SoC yet.


Considering two types of distributions, Bernoulli and Gaussian, frequentist approach based UCB algorithm remains the same while there are two variants of KL-UCB and Bayesian approach based TS algorithm, one each for Bernoulli and Gaussian distribution \cite{mab1,mab2,mab3,mab4,mab5,TAC1987_AsyptoticallyEfficient_AnantharamWalrand}. When arm distribution is known, we can select the appropriate algorithm and both guarantee lower regret than the UCB algorithm \cite{mab3}. 
When arm distribution is unknown, the challenge is to decide the correct variant of the KLUCB/TS algorithm, and error in algorithm selection leads to significant degradation in performance. Specifically, the use of the Bernoulli variant of the KLUCB/TS algorithm for Gaussian distribution or vice-versa leads to high regret compared to the UCB algorithm \cite{aggr1,aggr2}. This demands intelligence to select the appropriate algorithm in an unknown environment.


In this paper, we design and implement a reconfigurable and intelligent architecture for MAB algorithms (RI-MAB) that can learn and select an appropriate algorithm in an unknown environment so as to minimizes regret. The main contributions of this paper are summarized as below:
\begin{enumerate}
  \item We propose a synthesizable TS algorithm for Bernoulli distribution (BTS) by approximating the \textit{Beta} function via pseudo-random number generator-based approach and map it on Zynq SoC via hardware-software co-design. The architecture is optimized to reduce the computational complexity without compromising on the regret performance. 
 \item For an environment with unknown arm distribution, we propose RI-MAB architecture. Here, intelligence enables the identification of appropriate algorithm (UCB or BTS) for a given environment, and reconfigurability allows anytime on-the-fly switching between UCB and BTS algorithms via dynamic partial reconfiguration (DPR). The DPR on FPGA eliminates the need for parallel implementation of algorithms resulting in huge savings in resources and power consumption. 
 \item The functional correctness, resource requirement, power consumption, and execution time of the proposed BTS and RI-MAB architectures are analyzed for various arm distributions, word-length, and hardware-software co-design approaches.
 \item We also demonstrate the superiority of the proposed RI-MAB architecture over BTS and UCB only architectures.
\end{enumerate}

The rest of the paper is organized as follows. 
The MAB problem setup and a review of the relevant works are done in Section~\ref{Sec:RW} followed by the synthesizable TS algorithm in Section~\ref{Sec:TS}. The improved version of the TS algorithm and its architecture on SoC is presented in Section~\ref{Sec:ITS}. In Section~\ref{Sec:PITS}, in-depth performance analysis and comparison with the UCB algorithm is done. The RI-MAB algorithm and its architecture are discussed in Section~\ref{Sec:RIMAB} followed by its performance analysis in Section~\ref{Sec:PRIMAB}. Section~\ref{Sec:Conclusion} concludes the paper. Please refer to \cite{codes} for source codes and tutorial to reproduce results presented in this paper.

\section{MAB Algorithms and State-of-the-art Review}
\label{Sec:RW}
In this section, we discuss the MAB setup, review state-of-the-art MAB algorithms, and feasibility on the SoC. 

In MAB setup, each experiment consists of a horizon of $N$, $n \in \{1,2,...,N\}$ sequential slots and and the aim is to select the optimal arm from $K$, $k \in \{1,2,...,\}$ arms as often as possible. Let's denote the arm selected in slot $n$ as $I_n$, and the reward received from the selected arm $I_n$ in slot $n$ as $R_n$. The reward of an arm $k$ is generated from a distribution with mean, $\mu_k$. The mean rewards are unknown and the performance metric, regret, is given as \cite{mab1,mab2,mab3,mab4,mab5,TAC1987_AsyptoticallyEfficient_AnantharamWalrand}
\begin{equation}
    \label{regret}
    Regret = N*\mu^* - \sum_{n=1}^{N} R_n = N*\mu^* - \sum_{k=1}^{K} T_k *\mu_k 
\end{equation}
where $\mu^*$ is the mean reward of an optimal arm and $T_k$ is the number of times the arm $k$ selected in an experiment of horizon size $N$. 
 Note that the distribution of arm rewards is unknown but fixed over a horizon. In this paper, we focus on Bernoulli (reward $R_n$ is either 0 or 1) and Gaussian distribution (reward $R_n$ is between 0 and 1) though the discussion can be extended to Exponential and Poisson distributions. 

As discussed in Section I, UCB, KL-UCB, and TS algorithms are the popular regret-minimization MAB algorithms with logarithmic regret guarantees. In the case of the UCB and KL-UCB algorithms, the first phase is initialization where each arm is selected once in the first $K$ time slots. Thereafter, in each slot, quality factor (QF) $Q(k,n)$ is calculated for each arm. For UCB, the QF, $Q_u(k,n)$, is given by, \cite{mab1}

\begin{equation}
\label{qf_ucb}
    Q_u(k,n) = \frac{X(k,n)}{T(k,n)} + \sqrt{\frac{\alpha \log(n)}{T(k,n)}}
\end{equation}
where
\begin{equation}
\label{X}
    X(k,n) = X(k,n-1) + R_{n-1} \cdot \textbf{1}_{\{I_{n-1}==k\}} \quad \forall k
\end{equation}
\begin{equation}
\label{T}
    T(k,n) = T(k,n-1) + \textbf{1}_{\{I_{n-1}==k\}} \quad \forall k
\end{equation}
where $\textbf{1}_{cond}$ is an indicator function and it is equal to 1 (or 0) if the condition, $cond$ is TRUE (or FALSE). The parameter $X(k,n)$ is the total reward received using the arm $k$ which has been selected for $T(k,n)$ time slots in total $n$ time slots. The parameter, $\alpha$, is the exploration factor that quantifies the aggression by which the UCB algorithm explores all arms and theoretically, it lies between 0.5 and 2. Then, the arm with the highest QF is selected and it is denoted by, $I_n$.
\begin{equation}
\label{I}
    I_n = \argmax_k Q_u(:,n)
\end{equation}
After the arm $I_n$ is played, its parameters, $T$ and $X$, are updated using the received reward, $R_n$ as shown in Eq.~\ref{X} and Eq.~\ref{T}. The KL-UCB algorithm is similar to UCB except for the calculation of QF which is denoted as $Q_{kl}(k,n)$ \cite{mab2}. As shown in Eq.~\ref{qkl}, QF is computationally complex due to underlining optimization function \cite{mab2,mabarch1}. 
\begin{equation}
\label{qkl}
 Q_{kl}(k,n) = \max \bigg\{q \in \left [ 0,1 \right ],d\bigg(\frac{X(k,n)}{T(k,n)},q\bigg) \leq Y(k,n)\bigg\}
\end{equation}
where
\begin{equation}
 Y(k,n)=\frac{\log n+c\log(\log n)}{T(k,n)}
 \vspace{-0.15cm}
\end{equation}
\begin{equation}
\vspace{-0.1cm}
\label{kl}
d(p,q) = p\log\bigg(\frac{p}{q}\bigg) + (1-p) \log\bigg(\frac{1-p}{1-q}\bigg)
\end{equation}
The term, $d(p,q)$, in Eq.~\ref{kl} denotes the KL divergence between $p$ and $q$. The TS algorithm does not need an initialization phase and it uses \textit{Beta} function for QF calculation. It is discussed later in Section II. 

All these algorithms have been extended for various other settings. The multi-play setting is the same as above except that the aim is to identify the best $L$ arms instead of one arm \cite{multiplaymab1,multiplaymab2}. In a time-limited pure exploration setting, the aim is to identify the best arm within a given number of time-slots such that the regret incurred during these slots is not considered i.e. pure exploration phase \cite{pexp1,pexp2,pexp3}.  In a confidence-driven pure exploration setting, the aim is to identify the best arm with given confidence and in as few time slots as possible \cite{pexp1,pexp2,pexp3}. In a delayed and complex case, the reward of the selected arm in time slot $n$ is delayed by few time slots and such delay is not deterministic \cite{DCMAB1,DCMAB2,DCMAB3}. Also, the received reward might be a function of arms selected in multiple time slots instead of a separate reward for each slot \cite{DCMAB1,DCMAB2,DCMAB3}. Existing works mainly focus on the design and performance analysis of these algorithms while the focus of this work is on efficient mapping of MAB algorithms on the SoC. Since all these extensions are based on UCB/KLUCB/TS algorithms, an efficient implementation of these three algorithms is the first and important step towards the realization of all other algorithms on the SoC.


In \cite{mabarch2}, we discussed the mapping of the UCB algorithm and its extensions on Zynq SoC via a hardware-software co-design approach. In \cite{mabarch1} we proposed the modified KL-UCB algorithm by replacing optimization function in Eq.~\ref{qkl} with  finite-iteration based  synthesizable function. Though KL-UCB offers lower regret, the resource, latency, and power consumption of the KL-UCB is high compared to the UCB. To reduce the latency and power consumption without compromising on the regret performance, we proposed reconfigurable KL-UCB architecture that enables on-the-fly switch from KL-UCB to light-weight UCB after initial exploration  \cite{mabarch1}. In the proposed architecture, UCB QF calculation is accomplished using the KL-UCB QF blocks and hence, parallel implementation of two architectures is not needed. Since the TS is the most popular MAB algorithm, efficient mapping of the TS on SoC and performance analysis for different word-length is an important research problem. Furthermore, an intelligence to identify the appropriate algorithm in an unknown environment is critical to get optimal regret performance. The work presented in this paper offers innovative solutions to these challenges.





\section{Synthesizable Thompson Sampling Algorithm for Bernoulli Distribution (SBTS)}
\label{Sec:TS}
 The frequentist modeling-based UCB and KLUCB algorithms assume the mean reward of an arm is proportional to the average reward in repeated plays of a given experiment \cite{mab1,mab2,mab3,mab4,mab5,TAC1987_AsyptoticallyEfficient_AnantharamWalrand}. On the other hand, the Bayesian modeling-based TS algorithm assumes the mean reward of an arm is proportional to a degree of belief that the arm is optimal \cite{mab3}.  These beliefs are updated based on the observations from the environment via Baye’s rule that takes a prior belief as an argument and returns a posterior belief for a given likelihood. Since the arm statistics are unknown, the uncertainty about arm optimality is modeled as probabilities and the arm with the highest probability of being optimal under the posterior distribution is selected \cite{mab3}.
 
 
In the MAB setup, posterior belief becomes a prior in subsequent time slots, and the distributions which exhibit such behavior are known as conjugate prior. For example, Beta distribution is a conjugate prior for Bernoulli likelihood function  \cite{mab3}. Similarly, Gamma and Pareto distributions are a conjugate prior for Poisson and Gamma distributions, respectively  \cite{mab3}. Thus, the Bayesian approach needs to explicitly specify prior beliefs upfront in the form of the distribution and hence, each likelihood distributions have a specific variant of the TS algorithm. None of these TS variants are realized on the SoC yet and the proposed work on the implementation of the BTS algorithm on SoC is the first contribution in this direction.
 
The mapping of the BTS algorithm on the SoC consists of three steps: 1) Parameter update (Eq.~\ref{X} and Eq.~\ref{T}), 2) QF Calculation, 3) Arm selection (Eq.~\ref{I}). Since steps 1 and 3 are identical to UCB and KL-UCB algorithms, we request readers to refer to \cite{mabarch1,mabarch2} for in-depth understanding and implementation. Due to limited page constraints, the discussion is focused only on Step 3: QF calculation though our implementation and tutorials include all three steps. In the BTS algorithm, the QF for each arm, denoted by $Q_{ts}(k,n)$, is calculated by drawing the random sample from the Beta distribution with parameters, $\alpha = X(k,n)$ and $\beta = T(k,n)-X(k,n)$. For the Bernoulli distribution, $\alpha$ refers to a number of successes and $\beta$ refers to a number of failures. The probability distribution function (PDF), $y_{beta}$, of Beta distribution is given by \cite{mab3},
\begin{equation}
\label{beta_pdf}
    y_{beta} = f(x|\alpha,\beta) = \frac{1}{B(\alpha,\beta)}x^{\alpha-1}(1-x)^{\beta-1}I_{[0,1]}(x)
\end{equation}
 where $B(\cdot)$ is the Beta function. The indicator function $I_{[0,1]}(x)$ ensures that only values of x $\in (0,1)$ have nonzero probability. The QF calculation of the BTS algorithm involves two steps:\\
 1) Integration of the PDF, $y_{beta}$ given in Eq.~\ref{beta_pdf}, over $x$ to generate the cumulative distribution function (CDF), $F(x|\alpha,\beta)$ and computation of its inverse $F^{-1}(x|\alpha,\beta)$.
 \\
 2) Generation of a uniformly distributed random number $x$ and its substitution into inverse CDF. The value obtained is the random number sampled from the Beta distribution and it is considered as the QF of the arm.

The implementation of the above steps is computationally intensive and not well suited for hardware implementation due to the need for $gamma$ random generators followed by the division of random numbers. Please refer to the in-built Matlab function, $betarnd$ for more details. In the proposed approach, we approximate the $betarnd$ function using an alternative synthesizable function. In each time slot $n$, we generate $T(k,n)$ uniform random numbers for each arm $k$. These random numbers are sorted in ascending order and $X(k,n)^{th}$ random number in the sorted array is taken as the QF of the arm $k$. This approach needs the generation of random numbers in hardware and we implement a popular Mersenne Twister pseudo-random number generator (PRNG) due to its high throughput \cite{PRNG1,PRNG2,PRNG3}. We referred to it as a synthesizable BTS (SBTS) algorithm.

\begin{figure}[!h]
		\centering
	    \includegraphics[width=0.475\textwidth]{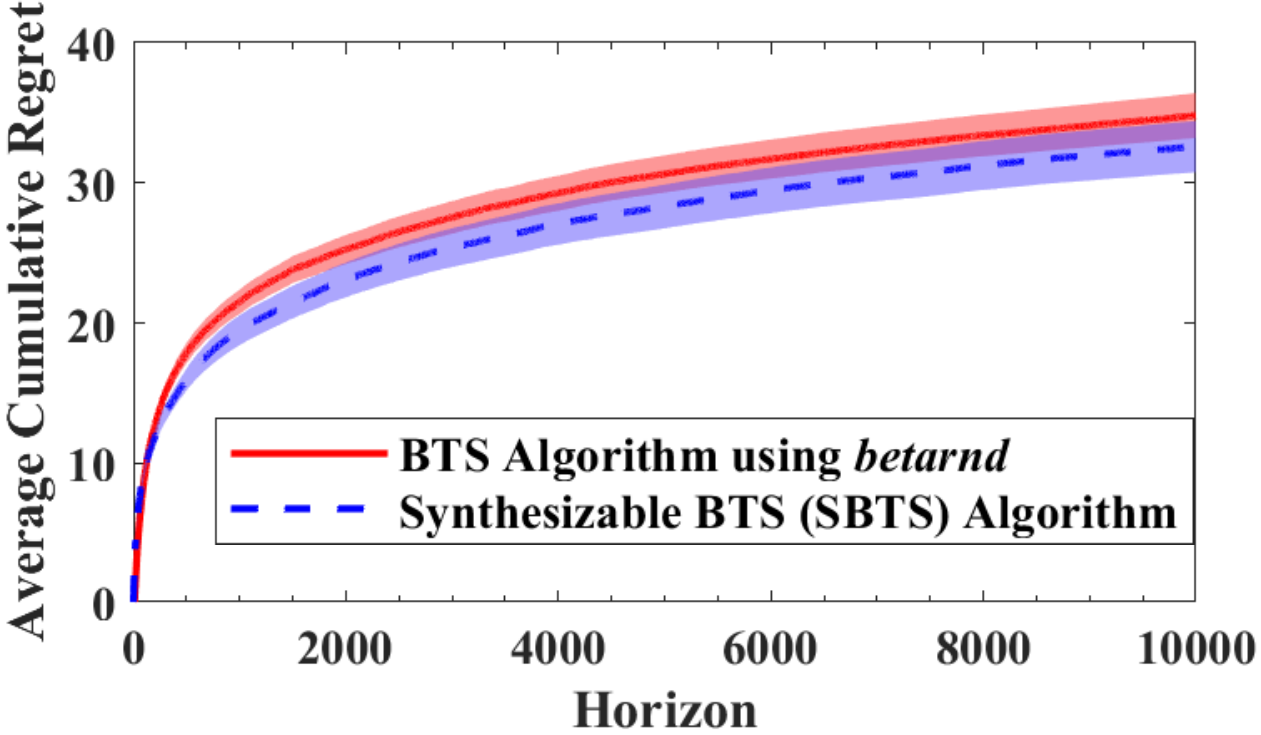}
		 \caption{ Cumulative regret averaged over 150 independent experiments for the BTS algorithm with QF using $berarnd$ and proposed SBTS algorithm.}
		\label{regretTS}
	\end{figure}

In Fig.~\ref{regretTS},  the functionality of the proposed SBTS algorithm is verified by comparing its regret performance with the BTS algorithm realized using the $betarnd$ function. We consider $K=6$, $N=10000$ and 150 experiments. In each experiment, arm statistics are chosen randomly and the plots in Fig.~\ref{regretTS} include the cumulative regret averaged over 150 experiments, and standard deviation (shown with a shaded region). We observed that the proposed approach selected the best arm on an average 9436  number of times ($\approx94\%$) compared to 9445  times ($\approx94\%$) in the $betarnd$ based BTS algorithm. The regret of both algorithms is nearly identical validating the correctness of the proposed QF calculation approach.

\begin{figure*}[!b]
		\centering
		\vspace{-0.2cm}
	    \includegraphics[width=0.9\textwidth]{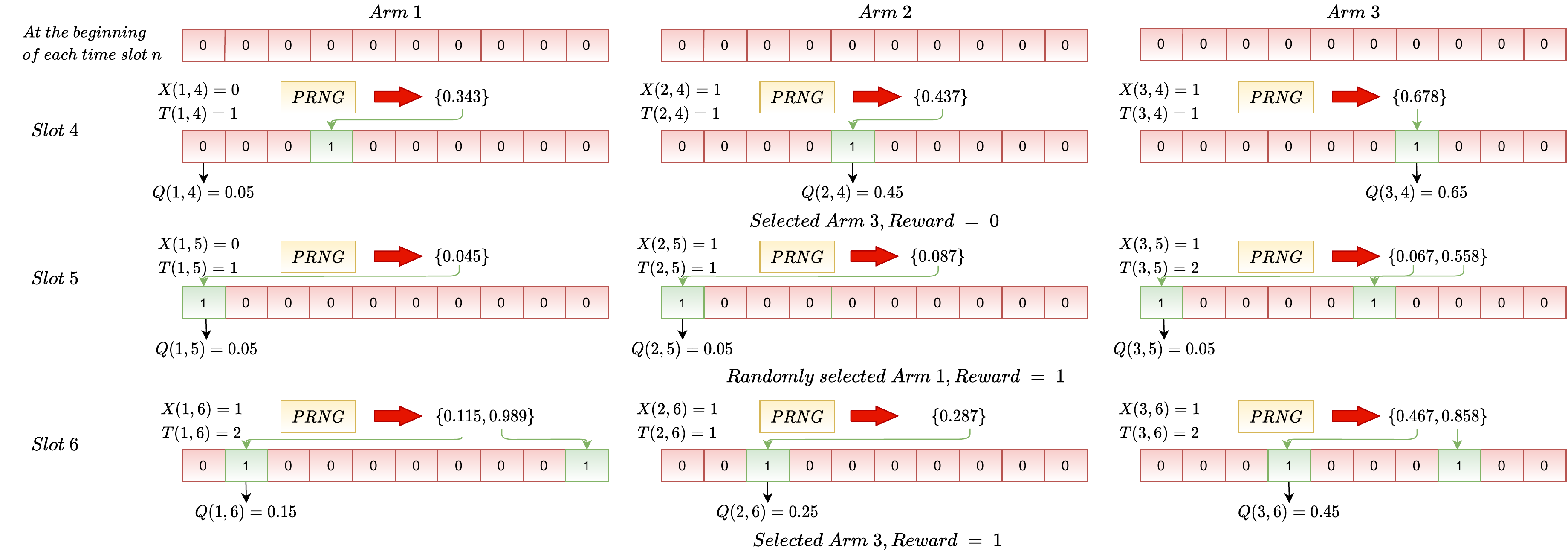}
		 \caption{Illustrative example demonstrating the functionality of the SBTS-ES algorithm.}
		\label{sbtses}
	\end{figure*}

\section{Improved SBTS Algorithm For Efficient Mapping on SoC}
\label{Sec:ITS}
The proposed SBTS QF calculation approach is synthesizable on the SoC but it suffers from two drawbacks: 
\begin{enumerate}
 \item In each time slot, a large number of random numbers need to be generated. For $K$ arms, we need to generate $\sum_{k=1}^{k=K}T(k,n)=n$ random numbers in each time slot $n$. This is followed by $K$ sorting operations, one for each arm. In the worst case (when n=N i.e. end of the horizon), we need to generate and sort $N$ random numbers in a slot. Thus, the time required to generate random numbers is not fixed and increases with $n$. This is not desirable for most applications.
\item Even if each random number is represented using fewer bits, say 8 bits, we need total storage of $N$ bytes (For example, 10 Kbytes when $N$=10000) which is not feasible due to cost and area constraints of the majority of the embedded applications.
\end{enumerate}

Ideally, MAB execution time should be fixed and as small as possible. The time taken by the MAB algorithm to select the arm affects the time available for subsequent tasks. For instance, in wireless radio, communication is time-slotted which means arm (channel) selection is followed by transmission in each time slot. The higher the time taken for channel selection, the lower is the time available for actual data communication resulting in lower throughput. In the BTS algorithm, the time required to calculate QF for all arms increases with time due to an increase in the number of random numbers and subsequent sorting tasks. Please refer to Section for the detailed resource requirement and execution time comparison. To overcome the above drawbacks of the SBTS algorithm, we present a further improvements to simplify the sorting operation and minimize the number of random number generation in each slot.

\subsection{SBTS-ES: SBTS Algorithm with Efficient Sorting}
In the MAB setup with $K$ arms, the SBTS algorithm involves sorting of $K$ arrays consisting of random numbers between 0 and 1. For arm $k$, array size is $T(k,n)$ with its maximum value as $N$. After sorting, random value at the $X(k,n)^{th}$ index of the sorted array is considered as QF of the arm. For accurate QF calculation, floating-point representation of random numbers is preferred which results in computationally complex sorting operation. 

In the proposed SBTS-ES algorithm, the sorting operation is simplified by grouping the random numbers in pre-defined ranges and keeping the track of number of random numbers generated in each range. For illustration, consider an array, $\beta_{k}$, of size $10$ such that $\beta_{k}(1)$ represents the number of random numbers out of $T(k,n)$ lies between 0 and 0.1, $\beta_{k}(2)$ represents the number of random numbers out of $T(k,n)$ lies between 0.1 and 0.2. In the same fashion, $\beta_{k}(10)$ represents the number of random numbers out of $T(k,n)$ lies between 0.9 and 1. For $k^{th}$ arm with $T(k,n)=4$ and $X(k,n)=2$, four random numbers are generated in time slot $n$. Lets assume these random numbers as \{0.342, 0.012, 0.753, 0.553\}. Then, $\beta_{k} = \{1,0,0,1,0,1,0,1,0,0\}$. The $X(k,n)^{th}$ non-zero value lies in the $\beta_{k}(3)$ and hence, QF of the arm is equal to the mean of $\beta_{k}(3)$ range i.e. $\frac{0.3+0.4}{2} = 0.35$. For random numbers as \{0.342, 0.012, 0.083, 0.553\}, we have $\beta_{k} = \{2,0,0,1,0,1,0,0,0,0\}$ and hence, the QF of the arm is equal to mean of $\beta_{k}(1)$ range i.e. $\frac{0+0.1}{2}= 0.05$. 

In Fig.~\ref{sbtses}, we consider $K=3$ arms. In the first $K$ time slots, each arm is selected once. In time slot 4, $T(k,n)$ random numbers are generated for $k^{th}$ arm. In Fig.~\ref{sbtses}, one random number is generated separately for each arm. After updating respective $\beta_k$, QF is calculated for each arm and the third arm is selected due to the highest QF. As shown in Fig.~\ref{sbtses}, received reward is 0 in time slot 4 and hence, only $T(3,5)$ is updated i.e. $T(3,5)=T(3,4)+1$. The rest of the parameters do not change. Note that $\beta_k$ of all arms is initialized to zero at the beginning of each time slot. In time slot 5, one random number is generated for the first two arms and two random numbers are generated for the third arm. The same process of QF calculation, arm selection, and parameter update are repeated in each time slot till the end of the horizon. The advantages of the proposed SBTS-ES are:

\begin{enumerate}
\item Instead of $\sum_{k=1}^{k=K}T(k,n)$ i.e. at most $N$ floating-point random numbers, the storage of only $|\beta|K$ integer numbers with word length of $\lceil \log_2{N} \rceil$ is needed.
 Here, $|\beta| = |\beta_1| =|\beta_2|= .. = |\beta_k|$. SBTS needs $32KN$ bits while SBTS-ES needs $|\beta|K\lceil \log_2{(N+1)} \rceil$. For $N \geq 100$ and $2\leq|\beta|\leq100$, SBTS memory requirement is at least 3 kilo Bytes (KB) higher than SBTS-ES. For $N > 1000$ and $N > 2000$, respectively, the difference is at least 50 KB and 80 KB, respectively. 
 \item In SBTS, sorting operations need multiple read and write memory operations for each random number which limits the execution time due to the limited number of memory ports. SBTS-ES needs only a single read and write operation per random number from a given index of $\beta$.
 \item In SBTS, each random number is compared with at most ($T(k,n)-1$) random numbers and hence, at most $(N-1)\sum_{k=1}^{k=K}T(k,n) \approx N^2$ floating point comparisons are done in each time slot. In SBTS-ES, only $2KN|\beta|$ comparisons are needed per slot. It is obvious to note that $N >> 2K|\beta|$ for sufficiently large horizon. For instance, for $K\leq 20$ and $|\beta| \leq 100$, SBTS-ES is superior for $N>4000$.  
\end{enumerate}

\subsection{SBTS-ESSR: SBTS-ES Algorithm with Single Random Number Sample}
In the SBTS-ES algorithm, total $\sum_{k=1}^{k=K}T(k,n)$ floating-point random numbers need to be generated in each time slot. This means total $N$ random numbers will be generated in the last time slot of the horizon. Generating such a huge number of random numbers is a time-consuming, memory-intensive, and inefficient approach. Since only one arm is selected in each time slot, the parameter $T(k,n)$ of all arms except the selected arm will remain unchanged. Though $T(k,n)$ random numbers are needed to calculate QF of an arm $k$, SBTS, and SBTS-ES algorithms discard previous generated random numbers. 
This is inefficient since instead of generating all random numbers in each slot, we can use the random numbers from previous slots as well. Furthermore, separate random number generators for each arm can be avoided.


In the proposed SBTS-ESSR approach, $\beta$ is not initialized at the beginning of each slot. In each time slot, a single random number is generated followed by updation of parameter $\beta$ for all arms. To incorporate a new random number, we discard any one of the entries in $\beta$ and update it with a newly generated random number. For illustrations, consider two arms with $\beta_1=\{0,1,0,0,1,1,0,0,1,1\}$ and $\beta_2=\{0,1,0,1,1,0,0,1,0,0\}$ in time slot 9. Then, $T(1,9)=5$ and $T(2,9)=4$. Assume $X(1,9)=2$ and $X(2,9)=3$. In the SBTS algorithm, 5 random numbers are generated for arm 1 followed by sorting and selection of $X(1,9)^{th}$ random number as the QF of the arm. The same process is repeated for arm 2 with 4 random numbers. In the SBTS-ES algorithm, existing $\beta$ is discarded and 9 random numbers are generated. Parameters, $\beta_1$, and $\beta_2$ are updated using these random numbers followed by QF calculation as discussed in Section. In the SBTS-ESSR algorithm, the first step is to randomly remove one sample from $\beta_1$ and $\beta_2$ instead of discarding them completely. Then, a single random number is generated which is used to update $\beta_1$ and $\beta_2$. After that, QF is selected using the same approach as in the SBTS-ES algorithm. It is important to note that SBTS-ESSR is a functional equivalent to SBTS-ES since the former generates $T(k,n)$ random numbers in a one-time slot while the latter uses ($T(k,n)-1$) random numbers generated in the previous time slots and only one random number is generated in the current time slot. Compared to SBTS-ES, SBTS-ESSR reduces the number of random number generations in each time slot as well as the number of comparisons by a factor $N$ i.e. from $2KN|\beta|$ to $2K|\beta|$. Furthermore, the execution time of the SBTS-ESSR algorithm is same in each time slot compared to SBTS and SBTS-ES algorithms where execution time in each slot increases with the increase in the index of the time slot.

The SBTS-ESSR algorithm is given in Algorithm 1. In the beginning, all elements of $X$ and $T$ are initialized to 1 assuming an initial uniform prior i.e. all arms have equal probability of being optimal. For clarity of notations, the subscript $n$ is removed in $X$ and $T$. In each time slot of the horizon, the QF of each arm is calculated (Line 2) and the arm with the highest QF is selected (Line 3). The selected arm is played and the algorithm receives the reward from the environment (Line 4). Based on the reward, parameters $X$ and $T$ are updated (Line 5).

      \setlength{\textfloatsep}{0pt}
        	\begin{algorithm}[!h]
			\caption{SBTS-ESSR Algorithm}
			\label{SBTSESSR}
			\textbf{Input:}$K, N$\\
			\textbf{Initialize:}~$X = [1]_{1\times K}, T = [1]_{1\times K}$\\
			\textbf{Output:}$Regret$
			\begin{algorithmic}[1]
				\For {$n=1,2,\cdots,N$}
				\State Calculate $Q_{ts}(:,n)$=\textbf{QF\_SBTS\_ESSR}($X$,$T$,$K$,$n$, $I_{n-1}$)
				\State Select and play arm, $I_n = \argmax_k Q_{ts}(:,n)$.
				\State Receive Reward, $R_n$ = \textbf{Reward}($I_n$).
				\State Update $X$ and $T$: $X(I_n)$ = $X(I_n)+R_n$, $T(I_n)$ = $T(I_n)+1$
            \EndFor
            \State Calculate regret using Eq.~\ref{regret}.
			\end{algorithmic}
		\end{algorithm}

The QF generation is explained using Subroutine 1. For a given $L=|\beta|$ and $K$, $\beta$ is a matrix where each column belongs to one arm. In the first time slot, $\beta$ is initialized in the same way as $T$ (Lines 1-3). Otherwise, $\beta$ is updated for the arm selected in the previous time slot (Lines 4-6) by generating a single random number. Then, one sample is removed from each column of $\beta$, and the corresponding row index is selected randomly (Lines 9 - 10). Next, a new random number is generated and $\beta$ of all arms is updated (Lines 11-13). Using updated $\beta$ and parameter $X$, the QF is calculated for each arm (Lines 14-15).

\begin{algorithm}[!h]
			\caption*{\textbf{Subroutine 1:} QF for SBTS-ESSR (QF\_SBTS\_ESSR)}
			\label{qfcalc}
			\textbf{Input:}$X$,$T$,$K$,$n$,$I_{n-1}$\\
			\textbf{Initialize:}~$L=$No. of divisions in the $[0,1]$ space i.e. $|\beta|$\\ 
			\textbf{Output:}~$Q_{1\times K}$
			\begin{algorithmic}[1]
			\If {$n == 1$}
				\State $\beta = [1]_{L\times K}$
			\Else
		\State Generate a random number, $p$ between 0 and 1. 
 			\State $\beta_{index} = l$ if $(l-1)/L \leq p < l/L$ where $l={1,2,..,L}$.
            	 \
            	 \State Update $\beta(\beta_{index},I_{n-1})$= $\beta(\beta_{index},I_{n-1})+1$. 
            	 	\EndIf
            	\For {$k=1,2,\cdots,K$}
            		\State Generate an integer random numbers, $s$, between 0 and $L$.
            		\State Update $\beta(s,:)$= $\beta(s,:)-1$. 
 			\State Generate a random number, $p$ between 0 and 1. 
 			\State $\beta_{index} = l$ if $(l-1)/L \leq p < l/L$ where $l={1,2,..,L}$.
            	 \
            	 \State Update $\beta(\beta_{index},:)$= $\beta(\beta_{index},:)+1$. 
 			\State Compute $Q_{index} :\argmin_l \sum_{l=1}^{L} \beta(:,k) \geq X(k)$
\State $Q(k)=(2Q_{index}-1)/(2L)$
            	\EndFor
			\end{algorithmic}
		\end{algorithm}
The environment generates the reward in each slot based on the selected arm and the corresponding process is given in Subroutine 2. Rewards are either 1 or 0 i.e. Bernoulli distribution. 		
\begin{algorithm}[!h]
			\caption*{\textbf{Subroutine 2:} Reward (Environment)}
			\label{rewardsr}
			\textbf{Input:}~$I_n$\\
			\textbf{Initialize:}~$\mu$ (Known to environment only)\\
			\textbf{Output:}~$R_n$
			\begin{algorithmic}[1]
			\State Generate random number, $p$ between 0 and 1. 
				\If {$p\leq \mu(I_n$)}
				\State $R_n$=1
				\Else
					\State $R_n$=0
				\EndIf
			\end{algorithmic}
		\end{algorithm}

\begin{figure}[!h]
		\centering
	    \includegraphics[width=0.475\textwidth]{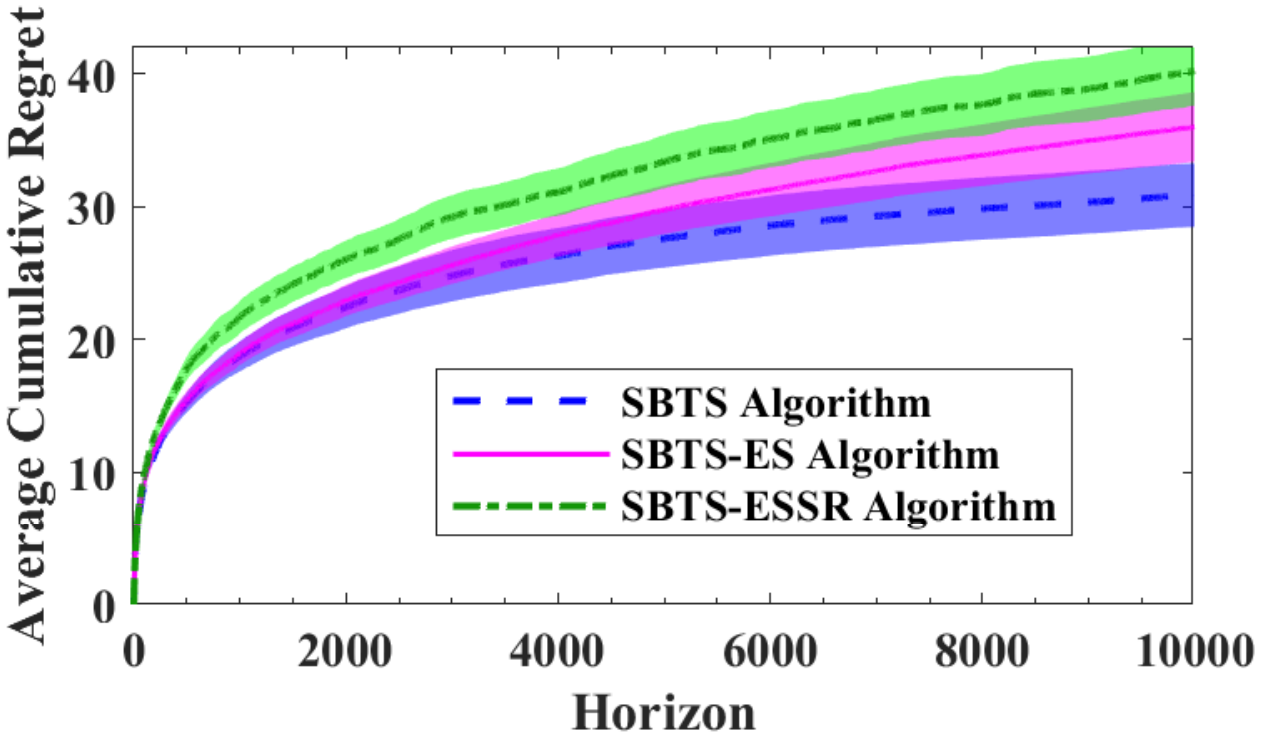}
		 \caption{Cumulative regret averaged over 150 independent experiments.}
		\label{regretTS1}
	\end{figure}

In Fig.~\ref{regretTS1}, we repeat the experiments similar to Fig.~\ref{regretTS} and compare the regret of the three proposed algorithms. As expected, the regret of the SBTS-ESSR is highest followed by SBTS-ES and SBTS. However, the difference in the regret is less than 12 for a horizon size of $N=10000$. On average, the best arm was selected 9421 (94\%), 9346 (93.5\%), 9271 (92\%) number of times. These results validate the functional correctness of the proposed algorithms in the MAB setup i.e. ability to identify and select the best arm as many times as possible.

\begin{figure}[!h]
     \includegraphics[width=0.475\textwidth]{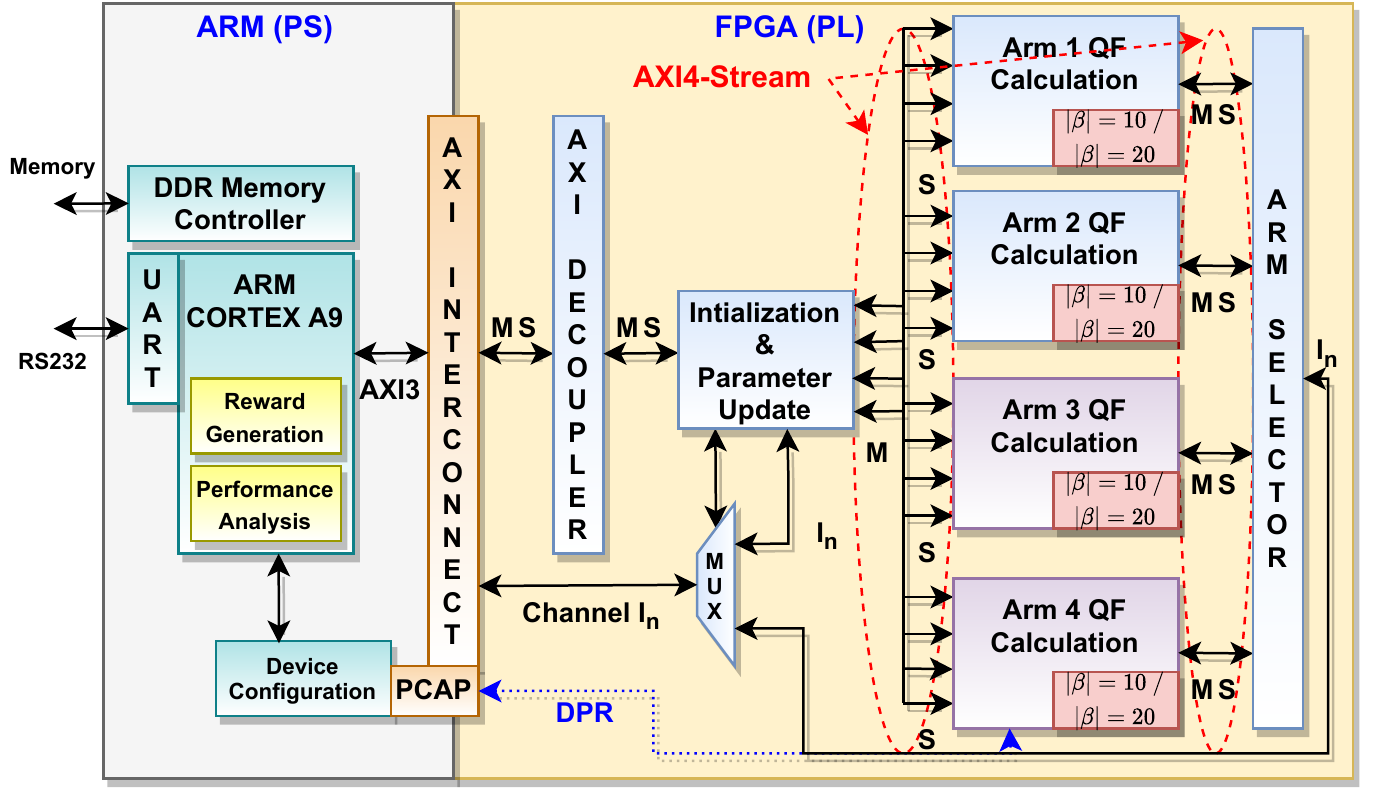}%
     \caption{ Proposed DPR-enabled architecture for SBTS and its extensions with reconfigurable $|\beta|$ and no. of arms.}
      \label{fig:ts_arch}
      \vspace{-0.2cm}
 \end{figure}

The proposed algorithms are mapped on the ZSoC platform and the corresponding architecture is shown in Fig.~\ref{fig:ts_arch}. The architecture is designed and implemented using Vivado 2019.1, Vivado High-Level Synthesis (HLS), Vivado Software Development Kit (SDK), and DPR toolbox.  The parameter update, QF calculation, and arm selection are realized in FPGA (PL) while reward generation is part of the ARM processor (PS). We have explored various other configurations via hardware-software co-design. Also, the WL of various blocks realized in FPGA is carefully chosen so as to optimize the resource utilization and power consumption without compromising on the regret performance. Corresponding results are presented in Section V. The section of the proposed architecture realized on FPGA is made reconfigurable via DPR. Specifically, the number of active arms, $K$, and $|\beta|$
can be dynamically configured via processor configuration access port (PCAP) using the partial bit-streams pre-loaded in the SD card \cite{DPR1,DPR2}. The required bitstreams are sent to the FPGA, through the bare-metal application deployed on the ARM processor, for reconfiguration using the device configuration (DevC) direct memory access (DMA). Please refer to \cite{codes} for source codes and tutorial explaining the building blocks of the proposed architectures.

\vspace{-0.2cm}
\section{Performance Analysis: SBTS Algorithms}
\label{Sec:PITS}
In this section, we verify the functional correctness of the proposed SBTS algorithms on Zynq SoC and compare its regret performance with state-of-the-art UCB algorithms for different WLs. The rewards are assumed to have Bernoulli distribution. All results are obtained after averaging over 100 different experiments to consider the non-deterministic nature of the online machine learning algorithms.
Later, the resource utilization, power consumption, and execution time of these algorithms are analyzed. MAB algorithms such as KL-UCB, UCB\_v, and UCB\_t are not considered since UCB offers regret which is close to these algorithms with significant savings in resources, power consumption and execution time \cite{mabarch1,mabarch2}. 


\subsection{Regret Comparison}
\begin{figure}[!h]
     \centering
     \includegraphics[width=0.5\textwidth]{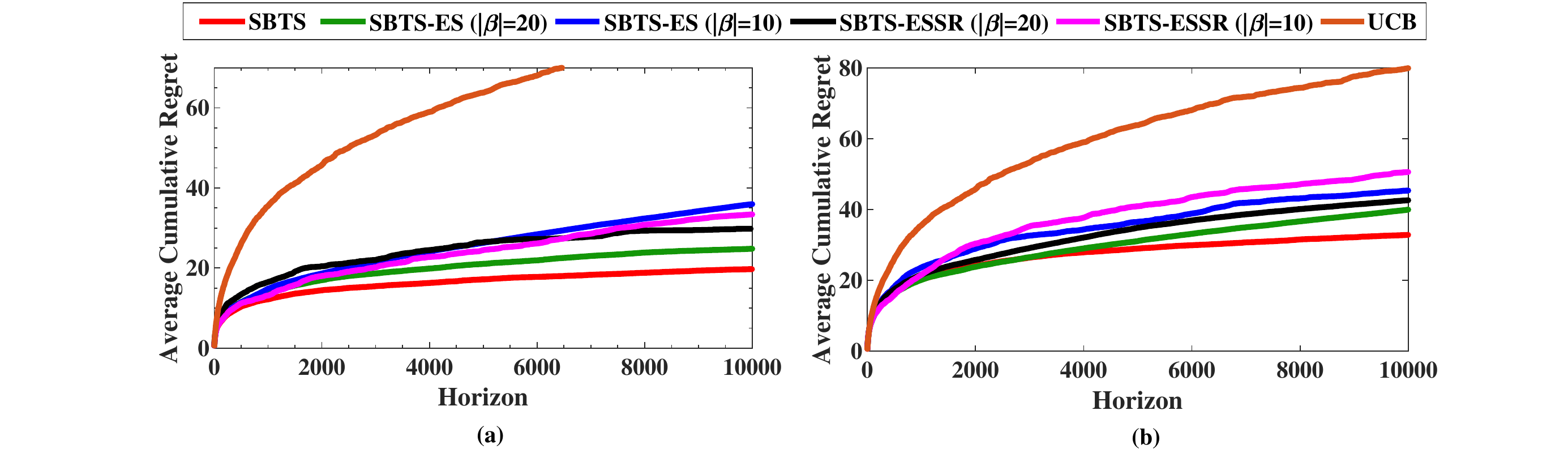}%
     \vspace{-0.2cm}
     \caption{ Average cumulative regret comparison for randomly generated arm distributions, (a) $K=4$, and (b) $K=8$.}
     
     \label{fig:reward_rnd_mu}
 \end{figure}

Similar to Fig.~\ref{regretTS} and Fig.~\ref{regretTS1}, we repeat the experiments for $K=4$ and $K=8$ for algorithms realized on ZSoC with single-precision floating-point (SP-FL) WL.  In Fig.~\ref{fig:reward_rnd_mu}, we consider four algorithms: 1) UCB, 2) SBTS, 3) SBTS-ES ($|\beta|=\{10,20\}$), and 4) SBTS-ESSR ($|\beta|=\{10,20\}$). It can be observed that the proposed SBTS algorithms offer significantly lower regret than UCB. This is expected since the TS algorithm has shown to outperform the UCB algorithm in analytical and simulation results. It can be observed that the regret of SBTS-ES and SBTS-ESSR algorithms decreases with an increase in $|\beta|$. This is because higher $|\beta|$ allows accurate calculation of QF leading to a reduction in the selection of sub-optimal arms.

\begin{table*}[!b]
\centering
\caption{Comparison of Resource Utilization and Power Consumption}
\label{table:TS_complexity}
\renewcommand{\arraystretch}{1.2}
\resizebox{\textwidth}{!}{
\begin{tabular}{|c|c|c|c|c|c|c|c|c|c|c|c|c|c|c|c|c|c|c|}
\hline
\multirow{3}{*}{\textbf{No.}} & \multirow{3}{*}{\textbf{Algorithm}}                                                                                & \multirow{3}{*}{\textbf{$|\beta|$}} & \multirow{3}{*}{\textbf{\begin{tabular}[c]{@{}c@{}}DPR/Velcro\\ Precision\end{tabular}}}                       & \multicolumn{3}{c|}{\multirow{2}{*}{\textbf{No. of LUTs}}}                                                                                                           & \multicolumn{3}{c|}{\multirow{2}{*}{\textbf{No. of FFs}}}                                                                                                            & \multicolumn{3}{c|}{\multirow{2}{*}{\textbf{No. of DSPs}}}                                                                                                   & \multicolumn{3}{c|}{\multirow{2}{*}{\textbf{No. of BRAMs}}}                                                                                                   & \multicolumn{3}{c|}{\multirow{2}{*}{\textbf{Dynamic Power (W)}}}                                                                                                      \\
                              &                                                                                                                    &                                                                                    &                                                                   & \multicolumn{3}{c|}{}                                                                                                                                                & \multicolumn{3}{c|}{}                                                                                                                                                & \multicolumn{3}{c|}{}                                                                                                                                        & \multicolumn{3}{c|}{}                                                                                                                                         & \multicolumn{3}{c|}{}                                                                                                                                                 \\ \cline{5-19} 
                              &                                                                                                                    &                                                                                    &                                                                   & \textbf{K=3}                                          & \textbf{K=4}                                          & \textbf{K=5}                                         & \textbf{K=3}                                          & \textbf{K=4}                                          & \textbf{K=5}                                         & \textbf{K=3}                                       & \textbf{K=4}                                       & \textbf{K=5}                                       & \textbf{K=3}                                        & \textbf{K=4}                                       & \textbf{K=5}                                       & \textbf{K=3}                                          & \textbf{K=4}                                          & \textbf{K=5}                                          \\ \hline \hline
\multirow{2}{*}{\textbf{1}}   & \multirow{2}{*}{\textbf{\begin{tabular}[c]{@{}c@{}}SBTS,\\ tunable $K \leq K_{max}$\end{tabular}}} & \multirow{2}{*}{NA}                                                                 & \textbf{\begin{tabular}[c]{@{}c@{}}DPR\\ SP-FP\end{tabular}}                                                 & \begin{tabular}[c]{@{}c@{}}14952\\ \textbf{-38\%}\end{tabular} & \begin{tabular}[c]{@{}c@{}}19488\\ \textbf{-19\%}\end{tabular} & 24024                                                & \begin{tabular}[c]{@{}c@{}}11804\\ \textbf{-38\%}\end{tabular} & \begin{tabular}[c]{@{}c@{}}15350\\ \textbf{-19\%}\end{tabular} & 18896                                                & \begin{tabular}[c]{@{}c@{}}18\\ \textbf{-44\%}\end{tabular} & \begin{tabular}[c]{@{}c@{}}24\\ \textbf{-25\%}\end{tabular} & 32                                                 & \begin{tabular}[c]{@{}c@{}}72\\ \textbf{-40\%}\end{tabular}  & \begin{tabular}[c]{@{}c@{}}96\\ \textbf{-20\%}\end{tabular} & 120                                                & \begin{tabular}[c]{@{}c@{}}0.258\\ \textbf{-39\%}\end{tabular} & \begin{tabular}[c]{@{}c@{}}0.339\\ \textbf{-20\%}\end{tabular} & 0.42                                                  \\ \cline{4-19} 
                              &                                                                                                                    &                                                                                    & \textbf{Velcro}                                                   & \multicolumn{3}{c|}{24024}                                                                                                                                           & \multicolumn{3}{c|}{18896}                                                                                                                                           & \multicolumn{3}{c|}{32}                                                                                                                                      & \multicolumn{3}{c|}{120}                                                                                                                                      & \multicolumn{3}{c|}{0.42}                                                                                                                                             \\ \hline \hline
\multirow{4}{*}{\textbf{2}}   & \multirow{4}{*}{\textbf{\begin{tabular}[c]{@{}c@{}}SBTS-ES,\\ tunable $K \leq K_{max}$\end{tabular}}} & \multirow{2}{*}{\textbf{10}}                                                       & \textbf{\begin{tabular}[c]{@{}c@{}}DPR\\ SP-FP\end{tabular}}                                                 & \begin{tabular}[c]{@{}c@{}}14634\\ \textbf{-38\%}\end{tabular} & \begin{tabular}[c]{@{}c@{}}19064\\ \textbf{-19\%}\end{tabular} & 23494                                                & \begin{tabular}[c]{@{}c@{}}11441\\ \textbf{-38\%}\end{tabular} & \begin{tabular}[c]{@{}c@{}}14866\\ \textbf{-19\%}\end{tabular} & 18291                                                & \begin{tabular}[c]{@{}c@{}}18\\ \textbf{-44\%}\end{tabular} & \begin{tabular}[c]{@{}c@{}}24\\ \textbf{-25\%}\end{tabular} & 32                                                 & \begin{tabular}[c]{@{}c@{}}24\\ \textbf{-40\%}\end{tabular}  & \begin{tabular}[c]{@{}c@{}}32\\ \textbf{-20\%}\end{tabular} & 40                                                 & \begin{tabular}[c]{@{}c@{}}0.178\\ \textbf{-39\%}\end{tabular} & \begin{tabular}[c]{@{}c@{}}0.234\\ \textbf{-20\%}\end{tabular} & 0.29                                                  \\ \cline{4-19} 
                              &                                                                                                                    &                                                                                    & \textbf{Velcro}                                                   & \multicolumn{3}{c|}{23494}                                                                                                                                           & \multicolumn{3}{c|}{18291}                                                                                                                                           & \multicolumn{3}{c|}{32}                                                                                                                                      & \multicolumn{3}{c|}{40}                                                                                                                                       & \multicolumn{3}{c|}{0.29}                                                                                                                                             \\ \cline{3-19} 
                              &                                                                                                                    & \multirow{2}{*}{\textbf{20}}                                                       & \textbf{\begin{tabular}[c]{@{}c@{}}DPR\\ SP-FP\end{tabular}}                                                 & \begin{tabular}[c]{@{}c@{}}20196\\ \textbf{-39\%}\end{tabular} & \begin{tabular}[c]{@{}c@{}}26480\\ \textbf{-20\%}\end{tabular} & 32764                                                & \begin{tabular}[c]{@{}c@{}}13488\\ \textbf{-38\%}\end{tabular} & \begin{tabular}[c]{@{}c@{}}17542\\ \textbf{-19\%}\end{tabular} & 21636                                                & \begin{tabular}[c]{@{}c@{}}18\\ \textbf{-44\%}\end{tabular} & \begin{tabular}[c]{@{}c@{}}24\\ \textbf{-25\%}\end{tabular} & 32                                                 & \begin{tabular}[c]{@{}c@{}}24\\ \textbf{-40\%}\end{tabular}  & \begin{tabular}[c]{@{}c@{}}32\\ \textbf{-20\%}\end{tabular} & 40                                                 & \begin{tabular}[c]{@{}c@{}}0.198\\ \textbf{-39\%}\end{tabular} & \begin{tabular}[c]{@{}c@{}}0.261\\ \textbf{-20\%}\end{tabular} & 0.324                                                 \\ \cline{4-19} 
                              &                                                                                                                    &                                                                                    & \textbf{Velcro}                                                   & \multicolumn{3}{c|}{32764}                                                                                                                                           & \multicolumn{3}{c|}{21636}                                                                                                                                           & \multicolumn{3}{c|}{32}                                                                                                                                      & \multicolumn{3}{c|}{40}                                                                                                                                       & \multicolumn{3}{c|}{0.324}                                                                                                                                            \\ \hline \hline
\multirow{6}{*}{\textbf{3}}   & \multirow{6}{*}{\textbf{\begin{tabular}[c]{@{}c@{}} SBTS-ESSR,\\ tunable $K \leq K_{max}$\end{tabular}}} & \multirow{2}{*}{\textbf{10}}                                                       & \textbf{\begin{tabular}[c]{@{}c@{}}DPR\\ SP-FP\end{tabular}}                                                 & \begin{tabular}[c]{@{}c@{}}7386\\ \textbf{-36\%}\end{tabular}  & \begin{tabular}[c]{@{}c@{}}9400\\ \textbf{-18\%}\end{tabular}  & 11414                                                & \begin{tabular}[c]{@{}c@{}}4631\\ \textbf{-34\%}\end{tabular}  & \begin{tabular}[c]{@{}c@{}}5786\\ \textbf{-17\%}\end{tabular}  & 6941                                                 & \begin{tabular}[c]{@{}c@{}}18\\ \textbf{-44\%}\end{tabular} & \begin{tabular}[c]{@{}c@{}}24\\ \textbf{-25\%}\end{tabular} & 32                                                 & \begin{tabular}[c]{@{}c@{}}27\\ \textbf{-40\%}\end{tabular}  & \begin{tabular}[c]{@{}c@{}}36\\ \textbf{-20\%}\end{tabular} & 45                                                 & \begin{tabular}[c]{@{}c@{}}0.046\\ \textbf{-40\%}\end{tabular} & \begin{tabular}[c]{@{}c@{}}0.061\\ \textbf{-20\%}\end{tabular} & 0.076                                                 \\ \cline{4-19} 
                              &                                                                                                                    &                                                                                    & \textbf{Velcro}                                                   & \multicolumn{3}{c|}{11414}                                                                                                                                           & \multicolumn{3}{c|}{6941}                                                                                                                                            & \multicolumn{3}{c|}{32}                                                                                                                                      & \multicolumn{3}{c|}{45}                                                                                                                                       & \multicolumn{3}{c|}{0.076}                                                                                                                                            \\ \cline{3-19} 
                              &                                                                                                                    & \multirow{4}{*}{\textbf{20}}                                                       & \textbf{\begin{tabular}[c]{@{}c@{}}DPR\\ SP-FP\end{tabular}} & \begin{tabular}[c]{@{}c@{}}7041\\ \textbf{-36\%}\end{tabular}  & \begin{tabular}[c]{@{}c@{}}8940\\ \textbf{-18\%}\end{tabular}  & 10839                                                & \begin{tabular}[c]{@{}c@{}}4811\\ \textbf{-34\%}\end{tabular}  & \begin{tabular}[c]{@{}c@{}}6026\\ \textbf{-17\%}\end{tabular}  & 7241                                                 & \begin{tabular}[c]{@{}c@{}}18\\ \textbf{-44\%}\end{tabular} & \begin{tabular}[c]{@{}c@{}}24\\ \textbf{-25\%}\end{tabular} & 32                                                 & \begin{tabular}[c]{@{}c@{}}27\\ \textbf{-40\%}\end{tabular}  & \begin{tabular}[c]{@{}c@{}}36\\ \textbf{-20\%}\end{tabular} & 45                                                 & \begin{tabular}[c]{@{}c@{}}0.096\\ \textbf{-40\%}\end{tabular} & \begin{tabular}[c]{@{}c@{}}0.127\\ \textbf{-20\%}\end{tabular} & 0.158                                                 \\ \cline{4-19} 
                              &                                                                                                                    &                                                                                    & \textbf{\begin{tabular}[c]{@{}c@{}}DPR\\ WL=27\end{tabular}} & \begin{tabular}[c]{@{}c@{}}1857\\ \textbf{-83\%}\end{tabular}  & \begin{tabular}[c]{@{}c@{}}2169\\ \textbf{-80\%}\end{tabular}  & \begin{tabular}[c]{@{}c@{}}2481\\ \textbf{-78\%}\end{tabular} & \begin{tabular}[c]{@{}c@{}}1792\\ \textbf{-76\%}\end{tabular}  & \begin{tabular}[c]{@{}c@{}}2110\\ \textbf{-71\%}\end{tabular}  & \begin{tabular}[c]{@{}c@{}}2428\\ \textbf{-67\%}\end{tabular} & \begin{tabular}[c]{@{}c@{}}0\\ \textbf{-100\%}\end{tabular} & \begin{tabular}[c]{@{}c@{}}0\\ \textbf{-100\%}\end{tabular} & \begin{tabular}[c]{@{}c@{}}0\\ \textbf{-100\%}\end{tabular} & \begin{tabular}[c]{@{}c@{}}3\\ \textbf{-94\%}\end{tabular}   & \begin{tabular}[c]{@{}c@{}}4\\ \textbf{-92\%}\end{tabular}  & \begin{tabular}[c]{@{}c@{}}5\\ \textbf{-89\%}\end{tabular}  & \begin{tabular}[c]{@{}c@{}}0.010\\ \textbf{-94\%}\end{tabular} & \begin{tabular}[c]{@{}c@{}}0.012\\ \textbf{-93\%}\end{tabular} & \begin{tabular}[c]{@{}c@{}}0.014\\ \textbf{-92\%}\end{tabular} \\ \cline{4-19} 
                              &                                                                                                                    &                                                                                    & \textbf{\begin{tabular}[c]{@{}c@{}}DPR\\ WL=11\end{tabular}} & \begin{tabular}[c]{@{}c@{}}1864\\ \textbf{-83\%}\end{tabular}  & \begin{tabular}[c]{@{}c@{}}2167\\ \textbf{-80\%}\end{tabular}  & \begin{tabular}[c]{@{}c@{}}2470\\ \textbf{-78\%}\end{tabular} & \begin{tabular}[c]{@{}c@{}}1792\\ \textbf{-76\%}\end{tabular}  & \begin{tabular}[c]{@{}c@{}}2110\\ \textbf{-71\%}\end{tabular}  & \begin{tabular}[c]{@{}c@{}}2428\\ \textbf{-67\%}\end{tabular} & \begin{tabular}[c]{@{}c@{}}0\\ \textbf{-100\%}\end{tabular} & \begin{tabular}[c]{@{}c@{}}0\\ \textbf{-100\%}\end{tabular} & \begin{tabular}[c]{@{}c@{}}0\\ \textbf{-100\%}\end{tabular} & \begin{tabular}[c]{@{}c@{}}3\\ \textbf{-94\%}\end{tabular}   & \begin{tabular}[c]{@{}c@{}}4\\ \textbf{-92\%}\end{tabular}  & \begin{tabular}[c]{@{}c@{}}5\\ \textbf{-89\%}\end{tabular}  & \begin{tabular}[c]{@{}c@{}}0.011\\ \textbf{-93\%}\end{tabular} & \begin{tabular}[c]{@{}c@{}}0.013\\ \textbf{-92\%}\end{tabular} & \begin{tabular}[c]{@{}c@{}}0.015\\ \textbf{-91\%}\end{tabular} \\ \cline{4-19} 
                              &                                                                                                                    &                                                                                    & \textbf{Velcro}                                                   & \multicolumn{3}{c|}{10839}                                                                                                                                           & \multicolumn{3}{c|}{7241}                                                                                                                                            & \multicolumn{3}{c|}{32}                                                                                                                                      & \multicolumn{3}{c|}{45}                                                                                                                                       & \multicolumn{3}{c|}{0.158}                                                                                                                                            \\ \hline \hline
\multirow{4}{*}{\textbf{4}}   & \multirow{4}{*}{\textbf{\begin{tabular}[c]{@{}c@{}}UCB,\\ tunable $K \leq K_{max}$\end{tabular}}}                                                          & \multirow{4}{*}{NA}                                                                 & \textbf{\begin{tabular}[c]{@{}c@{}}DPR\\ SP-FP\end{tabular}} & \begin{tabular}[c]{@{}c@{}}7664\\ \textbf{-37\%}\end{tabular}  & \begin{tabular}[c]{@{}c@{}}9866\\ \textbf{-19\%}\end{tabular}  & 12068                                                & \begin{tabular}[c]{@{}c@{}}7060\\ \textbf{-37\%}\end{tabular}  & \begin{tabular}[c]{@{}c@{}}9093\\ \textbf{-19\%}\end{tabular}  & 11126                                                & \begin{tabular}[c]{@{}c@{}}51\\ \textbf{-40\%}\end{tabular} & \begin{tabular}[c]{@{}c@{}}68\\ \textbf{-20\%}\end{tabular} & 85                                                 & \begin{tabular}[c]{@{}c@{}}4.5\\ \textbf{-40\%}\end{tabular} & \begin{tabular}[c]{@{}c@{}}6\\ \textbf{-20\%}\end{tabular}  & 7.5                                                & \begin{tabular}[c]{@{}c@{}}0.127\\ \textbf{-38\%}\end{tabular} & \begin{tabular}[c]{@{}c@{}}0.165\\ \textbf{-19\%}\end{tabular} & 0.203                                                 \\ \cline{4-19} 
                              &                                                                                                                    &                                                                                    & \textbf{\begin{tabular}[c]{@{}c@{}}DPR\\ WL=27\end{tabular}} & \begin{tabular}[c]{@{}c@{}}5761\\ \textbf{-53\%}\end{tabular}  & \begin{tabular}[c]{@{}c@{}}7375\\ \textbf{-39\%}\end{tabular}  & \begin{tabular}[c]{@{}c@{}}8989\\ \textbf{-26\%}\end{tabular} & \begin{tabular}[c]{@{}c@{}}4390\\ \textbf{-61\%}\end{tabular}  & \begin{tabular}[c]{@{}c@{}}5574\\ \textbf{-50\%}\end{tabular}  & \begin{tabular}[c]{@{}c@{}}6758\\ \textbf{-40\%}\end{tabular} & \begin{tabular}[c]{@{}c@{}}0\\ \textbf{-100\%}\end{tabular} & \begin{tabular}[c]{@{}c@{}}0\\ \textbf{-100\%}\end{tabular} & \begin{tabular}[c]{@{}c@{}}0\\ \textbf{-100\%}\end{tabular} & \begin{tabular}[c]{@{}c@{}}0\\ \textbf{-100\%}\end{tabular}  & \begin{tabular}[c]{@{}c@{}}0\\ \textbf{-100\%}\end{tabular} & \begin{tabular}[c]{@{}c@{}}0\\ \textbf{-100\%}\end{tabular} & \begin{tabular}[c]{@{}c@{}}0.019\\ \textbf{-91\%}\end{tabular} & \begin{tabular}[c]{@{}c@{}}0.023\\ \textbf{-89\%}\end{tabular} & \begin{tabular}[c]{@{}c@{}}0.027\\ \textbf{-87\%}\end{tabular} \\ \cline{4-19} 
                              &                                                                                                                    &                                                                                    & \textbf{\begin{tabular}[c]{@{}c@{}}DPR\\ WL=11\end{tabular}} & \begin{tabular}[c]{@{}c@{}}2425\\ \textbf{-80\%}\end{tabular}  & \begin{tabular}[c]{@{}c@{}}2927\\ \textbf{-76\%}\end{tabular}  & \begin{tabular}[c]{@{}c@{}}3429\\ \textbf{-72\%}\end{tabular} & \begin{tabular}[c]{@{}c@{}}2443\\ \textbf{-79\%}\end{tabular}  & \begin{tabular}[c]{@{}c@{}}2978\\ \textbf{-74\%}\end{tabular}  & \begin{tabular}[c]{@{}c@{}}3513\\ \textbf{-69\%}\end{tabular} & \begin{tabular}[c]{@{}c@{}}0\\ \textbf{-100\%}\end{tabular} & \begin{tabular}[c]{@{}c@{}}0\\ \textbf{-100\%}\end{tabular} & \begin{tabular}[c]{@{}c@{}}0\\ \textbf{-100\%}\end{tabular} & \begin{tabular}[c]{@{}c@{}}0\\ \textbf{-100\%}\end{tabular}  & \begin{tabular}[c]{@{}c@{}}0\\ \textbf{-100\%}\end{tabular} & \begin{tabular}[c]{@{}c@{}}0\\ \textbf{-100\%}\end{tabular} & \begin{tabular}[c]{@{}c@{}}0.013\\ \textbf{-94\%}\end{tabular} & \begin{tabular}[c]{@{}c@{}}0.015\\ \textbf{-93\%}\end{tabular} & \begin{tabular}[c]{@{}c@{}}0.017\\ \textbf{-92\%}\end{tabular} \\ \cline{4-19} 
                              &                                                                                                                    &                                                                                    & \textbf{Velcro}                                                   & \multicolumn{3}{c|}{12068}                                                                                                                                           & \multicolumn{3}{c|}{11126}                                                                                                                                           & \multicolumn{3}{c|}{85}                                                                                                                                      & \multicolumn{3}{c|}{7.5}                                                                                                                                      & \multicolumn{3}{c|}{0.203}                                                                                                                                            \\ \hline

\end{tabular}
}
\end{table*}

\begin{figure}[!h]
     \centering
   \includegraphics[width=0.5\textwidth]{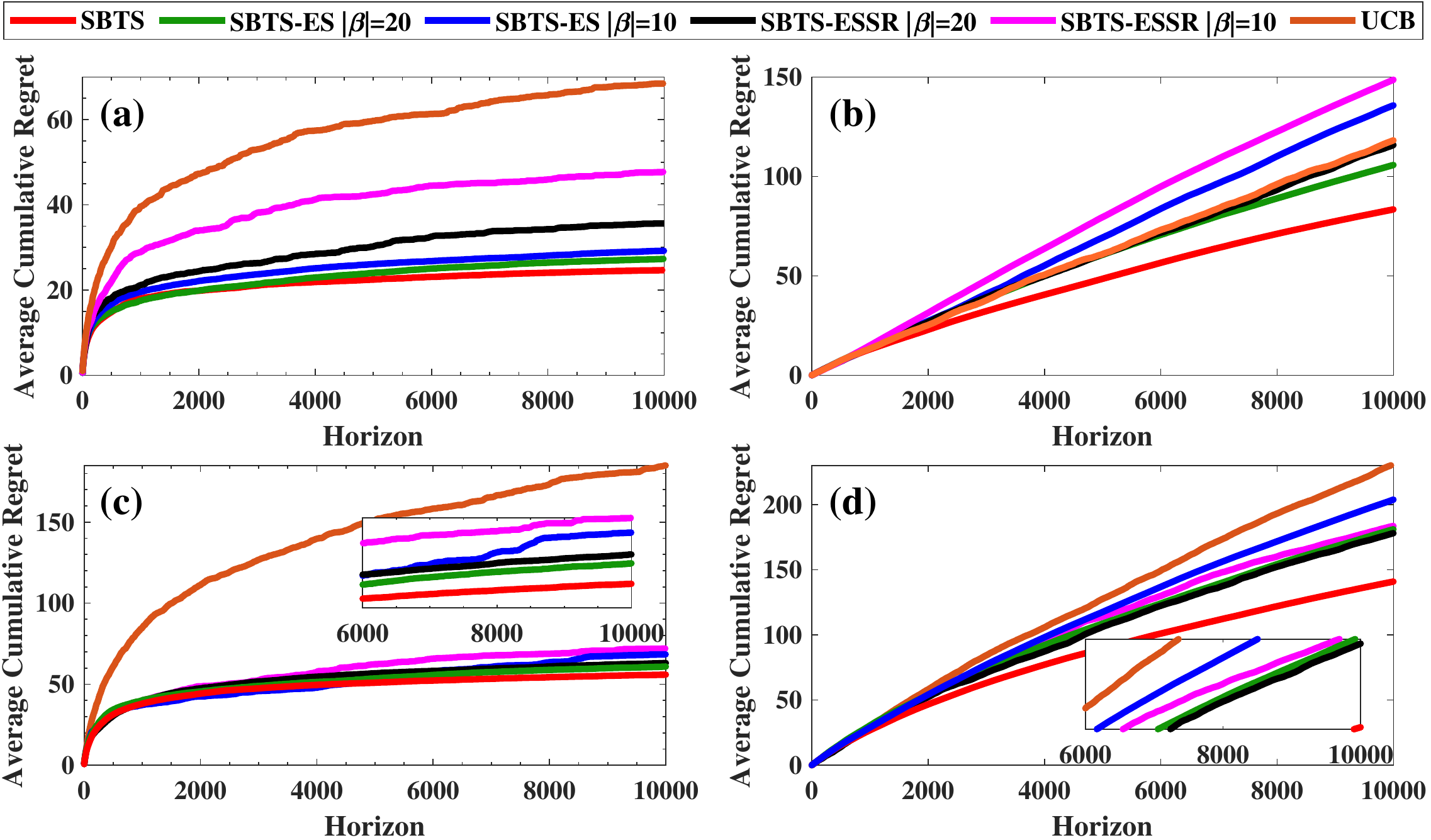}%
         \label{rew1}
     \caption{ Average cumulative regret of different algorithms for (a) $\mu_1$, (b) $\mu_2$,  (c) $\mu_3$, and (d) $\mu_4$.}
     \label{fig:reward_fix_mu}
 \end{figure}

\begin{figure}[!h]
     \centering
     \includegraphics[width=0.45\textwidth]{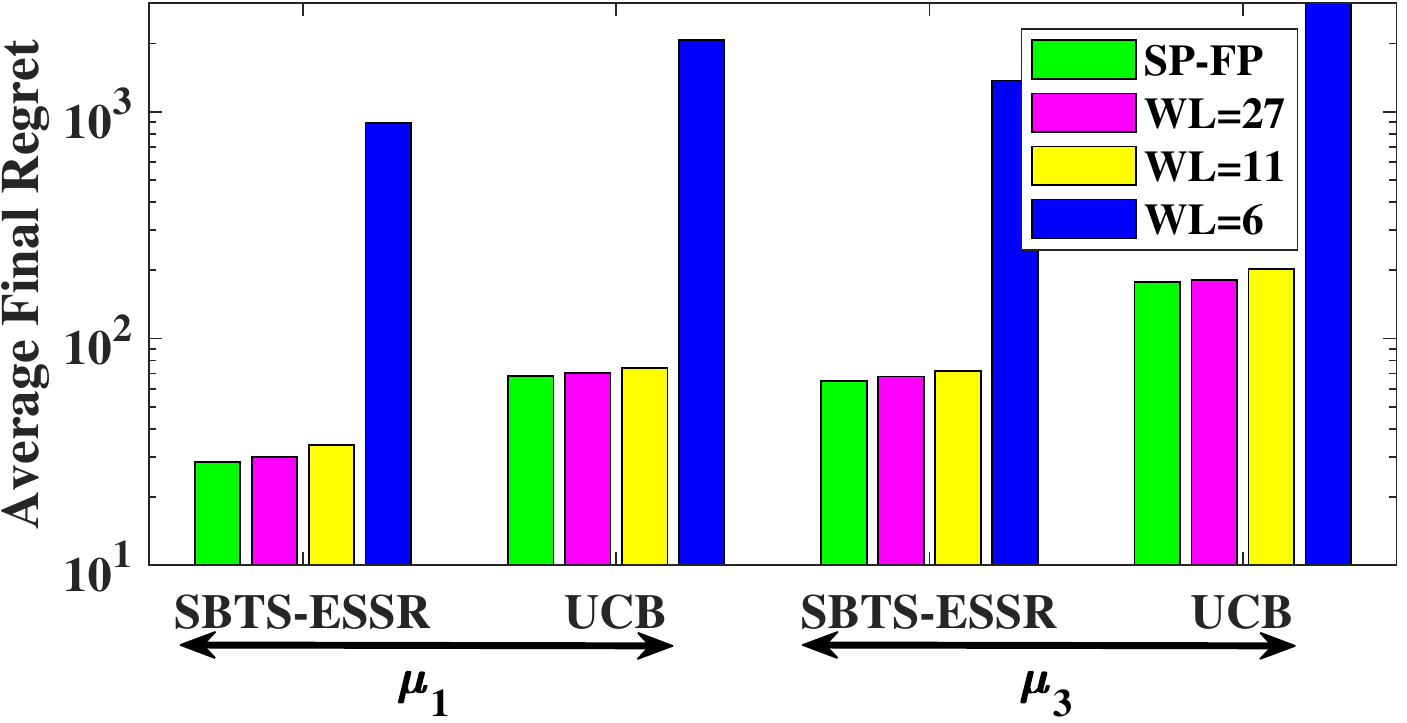}
     \caption{Average regret of SBTS-ESSR and UCB algorithms with different WLs for (a) $\mu_1$, and (b) $\mu_3$.}
     \label{fig:reward_diff_wl}
 \end{figure}

Next, we analyze the regret performance for four cases of carefully selected arm distributions.\\
1) $K=4$, $\mu_1$ = \{0.1, 0.3, 0.5, \textbf{0.7}\}\\
2) $K=4$, $\mu_2$ = \{\textbf{0.54},0.53, 0.52, 0.51\} \\
3) $K=8$, $\mu_3$ = \{0.1, 0.5, \textbf{0.8}, 0.7, 0.4, 0.2,0.6,0.3\} \\
4) $K=8$, $\mu_4$ = \{0.21, 0.22, 0.26,\textbf{0.28}, 0.24, 0.25, 0.27, 0.23\} \\
The optimal arm in each case is highlighted in bold font. Compared to $\mu_1$ and $\mu_3$, the difference between the arm statistics is small in $\mu_2$ and $\mu_4$. This makes the learning and identification of the optimal arm challenging. As shown in Fig.~\ref{fig:reward_fix_mu}, the regret of the SBTS, SBTS-ES, and SBTS-ESSR algorithms is lower than that of the UCB algorithm for all arm distributions.  In each case, it is verified that the optimal arm is chosen the highest number of times by all algorithms. The appropriate selection of $|\beta|$ is important as it affects the precision of QF selection. This results in multiple arms with identical QF values and hence, frequent selection of sub-optimal arms. For instance, $|\beta|=10$ is not sufficient for $\mu_2$ as it offers high regret as shown in Fig.~\ref{fig:reward_fix_mu} (d).Based on extensive performance analysis, $15\leq|\beta|\leq 20$ leads to a higher number of selection of optimal arm i.e. lower regret and the gain in performance is not significant for $|\beta|>20$. The proposed architecture in Fig.~\ref{fig:ts_arch} allows on-the-fly selection of $|\beta|$ via DPR.
Next, we compare the effect of WL on the regret performance of the SBTS-ESSR ($|\beta|=20$) and UCB algorithms. In Fig~\ref{fig:reward_diff_wl},
we compare the regret of these algorithms at the end of the horizon of size, $N=10000$ for $\mu_1$ and $\mu_3$. We compare the regret for SP-FP and fixed-point implementations with a total WL of 27, 11, and 6 bits. In each case, the number of bits to represent integer and fractional parts are chosen carefully so as to minimize regret. It can be observed that the regret degrades with the decrease in WL. However, degradation is not significant till WL=11. For WL=6, regret is high and this happens due to insufficient bits to represent QF which in turn leads to the selection of sub-optimal arms. Thus, the selection of appropriate WL is important since lower WL leads to significant savings in resources but it should not come at the cost of regret performance.


\subsection{Complexity Comparison}


In Table~\ref{table:TS_complexity}, we compare the resource utilization (LUT, FFs, DSP and BRAM), and power consumption of four different architectures realized on ZSoC. These architectures correspond to SBTS, SBTS-ES, SBTS-ESSR, and UCB algorithms. Each architecture is made reconfigurable via DPR at the arm level i.e. the number of arms can be dynamically configured (i.e., the number of arms can be tuned to any values less than or equal to $K_{max}$.  Each architecture is realized with three different WLs: SP-FP, fixed point WL of 27, and 11 bits. The reconfigurable architecture is compared with non-reconfigurable Velcro approach-based architecture with fixed $K_{max}$ arms and SP-FP WL.

As shown in Table~\ref{table:TS_complexity}, resource utilization and power consumption of DPR based architecture depend on the number of active arms, $K$ compared to the Velcro approach, which corresponds to architecture with $K_{max}$ arms, i.e., all blocks are active all the time compared to dynamic activation and deactivation of arms in the DPR based architecture.
It can be observed that the SP-FP version of the DPR-based architecture offers around savings of 18-39\% in LUTs, 17-38\% in FFs, 25-44\% in DSP48Es, and 20-40\% in BRAM over the Velcro approach for $K < K_{max}$. Further, they offer a 19-40\% reduction in the dynamic power consumption over the Velcro approach for $K < K_{max}$. Using the fixed-point implementation with WL=27, one can achieve up to 83\%, 79\%, 100\%, 100\% savings in the consumption of LUTs, FFs, DSP48Es, and BRAM\_18Ks respectively for $K < K_{max}$. This can be achieved with almost identical performance as that of the SP-FP architecture. With WL=11, further improvement in savings of up can be achieved with a slight degradation in the regret. In terms of the dynamic power consumed, the architectures with fixed-point WLs offer up to 94\% savings. Among SBTS algorithms, the SBTS-ESSR algorithm offers significant savings in resource utilization as well as power consumption. Compared to the UCB algorithm, SBTS-ESSR is computationally efficient, offers lower regret and power consumption. This makes the proposed SBTS-ESSR a superior alternative to the state-of-the-art UCB algorithm for the environment with Bernoulli rewards.

Execution time of the algorithm is an importance-performance metric that depends on efficient implementation and underlining architecture. In Table~\ref{table:latency}, we consider three different configurations obtained by realizing the algorithm using: 1) PS and PL with optimal partitioning, 2) Only PS (ARM + NEON), 3) Only ARM. It can be observed that the first approach offers the lowest execution time due to the parallel execution of the QF function in PL compared to sequential PS execution. Furthermore, the gain improves as $K$ increases. Among various TS algorithms, SBTS-ESSR offers the lowest execution time as expected. In applications like wireless networks, MAB algorithms are realized in upper layers (MAC/Network) i.e. in ARM or other processors while the PHY is present in the SoC. The proposed architecture enables shifting of the MAB algorithms from MAC to PHY layers thereby resulting in an accelerator factor ranging from 465.6-776 for UCB and 2.5-33.6 for TS. For larger $K > 20$, the acceleration factor will be significantly higher. Between UCB and TS algorithms, UCB is faster on ZSoC due to hardware-friendly arithmetic operations compared to PRNG in SBTS but UCB incurs high regret. On the PS-only architectures (i.e. ARM and ARM+NEON), SBTS-ESSR offers the lowest execution time. 

\begin{table}[!h]
\caption{Comparison of Execution Time in Milliseconds}
\label{table:latency}
\renewcommand{\arraystretch}{1.3}
\resizebox{0.5\textwidth}{!}{%
\begin{tabular}{|c|c|c|c|c|c|c|c|c|c|c|}
\hline
\multirow{2}{*}{\textbf{No.}} & \multirow{2}{*}{\textbf{Algorithm}} & \multirow{2}{*}{\textbf{Nb}} & \multirow{2}{*}{\textbf{Precision}} & \textbf{\begin{tabular}[c]{@{}c@{}}ZSoC\\ (in ms)\end{tabular}} & \multicolumn{3}{c|}{\textbf{\begin{tabular}[c]{@{}c@{}}PS (ARM)\\ (in ms)\end{tabular}}} & \multicolumn{3}{c|}{\textbf{\begin{tabular}[c]{@{}c@{}}PS (ARM+NEON)\\ (in ms)\end{tabular}}} \\ \cline{5-11} 
                              &                                     &                              &                                     & \textbf{K=\{3,4,5\}}                                            & \textbf{K=3}                 & \textbf{K=4}                & \textbf{K=5}                & \textbf{K=3}                  & \textbf{K=4}                  & \textbf{K=5}                  \\ \hline \hline
\textbf{1}                    & \textbf{SBTS}                       & \textbf{NA}                  & \textbf{SP-FP}                      & 12504                                                           & 26758                        & 35697                       & 44580                       & 13060                         & 17389                         & 21746                         \\ \hline \hline
\multirow{2}{*}{\textbf{2}}   & \multirow{2}{*}{\textbf{SBTS-ES}}   & \textbf{20}                  & \multirow{2}{*}{\textbf{SP-FP}}     & 18008                                                           & 35454                        & 47267                       & 59079                       & 6791                          & 8981                          & 11212                         \\ \cline{3-3} \cline{5-11} 
                              &                                     & \textbf{10}                  &                                     & 10010                                                           & 25134                        & 33509                       & 41882                       & 6549                          & 8730                          & 10913                         \\ \hline \hline
\multirow{4}{*}{\textbf{3}}   & \multirow{4}{*}{\textbf{SBTS-ESSR}} & \multirow{3}{*}{\textbf{20}} & \textbf{SP-FP}                      & 3.9                                                             & \multirow{3}{*}{38}          & \multirow{3}{*}{46}         & \multirow{3}{*}{54}         & \multirow{3}{*}{15}           & \multirow{3}{*}{16}           & \multirow{3}{*}{18}           \\ \cline{4-5}
                              &                                     &                              & \textbf{WL=27}                      & 1.5                                                             &                              &                             &                             &                               &                               &                               \\ \cline{4-5}
                              &                                     &                              & \textbf{WL=11}                      & 1.5                                                             &                              &                             &                             &                               &                               &                               \\ \cline{3-11} 
                              &                                     & \textbf{10}                  & \textbf{SP-FP}                      & 2.7                                                             & 33                           & 40                          & 46                          & 11                            & 13                            & 14                            \\ \hline \hline
\multirow{3}{*}{\textbf{4}}   & \multirow{3}{*}{\textbf{UCB}}       & \multirow{3}{*}{\textbf{NA}} & \textbf{SP-FP}                      & 0.1                                                             & \multirow{3}{*}{29}          & \multirow{3}{*}{37}         & \multirow{3}{*}{47}         & \multirow{3}{*}{20}           & \multirow{3}{*}{26}           & \multirow{3}{*}{33}           \\ \cline{4-5}
                              &                                     &                              & \textbf{WL=27}                      & 0.1                                                             &                              &                             &                             &                               &                               &                               \\ \cline{4-5}
                              &                                     &                              & \textbf{WL=11}                      & 0.1                                                             &                              &                             &                             &                               &                               &                               \\ \hline
\end{tabular}
}
\end{table}

\section{Reconfigurable and Intelligent MAB (RI-MAB)}
\label{Sec:RIMAB}
 \begin{figure*}[!b]
     \centering
     \includegraphics[width=0.9\textwidth]{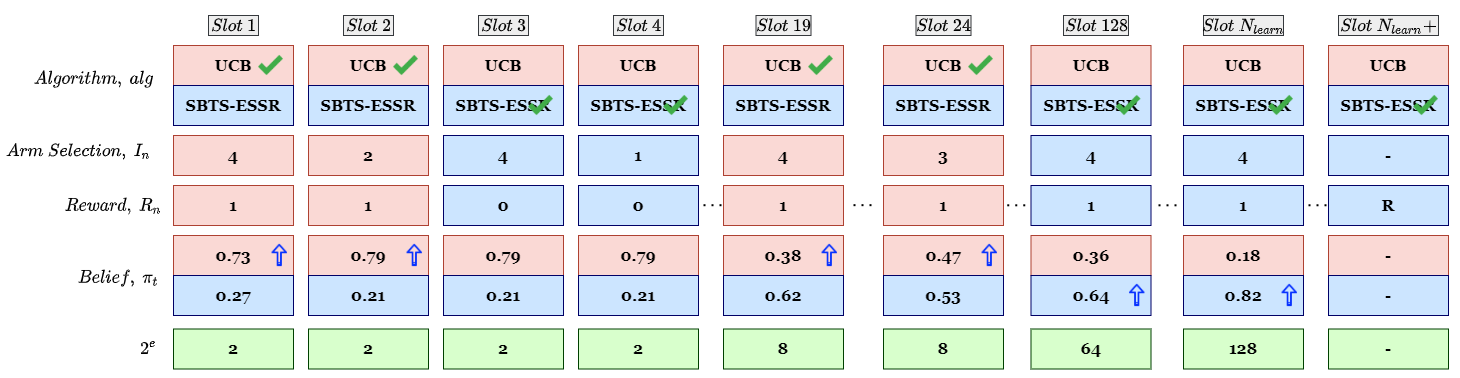}
     \caption{Working of the proposed RI-MAB algorithm in an example setting with $K=4$, $N=10000$, $N_{learn}=500$, $A=2$ and randomly generated arm distribution}
     \label{fig:agg_working}
 \end{figure*}
The SBTS-ESSR algorithm is well-suited only for arm rewards with Bernoulli distribution and hence, it may not outperform the UCB algorithm when arm distribution is unknown and random \cite{mab3}. For example, in wireless Radio, arm i.e. wireless channel distribution is usually Bernoulli at high signal-to-noise ratios (SNR) and Gaussian at medium and low SNRs. In IoT and robotics applications, the type of arm distribution is unknown. Such applications demand architecture which can learn, identify and deploy appropriate MAB algorithm.  

Very few works have addressed this problem and \cite{aggr1} is one of the recent works in which the aggregator algorithm selects the arm chosen by one of the candidate MAB algorithms in each time slot. The aim is to identify an optimal algorithm for a given unknown environment and this is achieved by learning from the past performance of various algorithms i.e algorithm exploration-exploitation trade-off in addition to arm exploitation-exploration trade-off. The major problem with \cite{aggr1} is the need for hardware implementation of all candidate MAB algorithms in parallel (referred here as Velcro MAB). Such architecture is area and power inefficient. The proposed RI-MAB algorithm augmented with DPR-based reconfigurable architecture overcomes this problem.


\begin{algorithm}[!b]
\caption{RI-MAB Algorithm}
\label{agg_algorithm}			
\begin{algorithmic}[1]
\State \textbf{Input:} $K$, $N$, $N_{learn}$, $A$
\State \textbf{Initialize:} $\pi_0 = U(\{1,2,..,A\}), e=1, alg=1, X = [1]_{1\times K}, T = [1]_{1\times K}, I_0=K, r=0$
\State \textbf{Output:} \textit{Regret}
\For{$n=1:1:N$}
    \If{$n\leq N_{learn}$}
       \State $I_n$ = \textbf{ARM\_SEL}($alg,n,I_{n-1},X,T,K$)
       \State Receive Reward, $R_n$ = \textbf{Reward}($I_n$) \Comment{Subroutine 2}
        \State Update the learning rate, $\eta_n=\sqrt{\log A/(n\times K)}$
        \State Compute the unbiased reward using the belief of the selected algorithm,
         $\hat{R}_{n}= \frac{R_n}{\pi_{n-1}(alg)}$
         \State Update the belief of the selected algorithm: $\pi_{n}(alg)=exp(\eta_{n}\hat{R}_{n})\times\pi_{n-1}(alg)$
         \State Normalize the belief of all algorithms: $\pi_{n}(i)=\pi_{n}(i)/\sum_{a=1}^2\pi_{n} (alg)$, $i\in\{1,2\}$
         \State {$r=r+1$}
        \If{{$r==2^e$} }
            \State {$r = 0$}
            \State $alg = (alg+1)$ \Comment{Switch Algorithm}
            \If {$alg>A$} 
            \State $e=e+1$\Comment{Increase epoch size}
            \State $alg=1$ \Comment{Reset to first algorithm}
            \EndIf
        \EndIf
    \Else \Comment{Select algorithm with higher belief}
    \State  $alg = \argmax \pi_{N_{learn}}-1$
    \State $I_n$ = \textbf{ARM\_SEL}($alg,n,I_{n-1},R_{n-1}$)
    \State Receive Reward, $R_n$ = \textbf{Reward}($I_n$) \Comment{Subroutine 2}
    \EndIf
    	\State Update $X$ and $T$: $X(I_{n})$ = $X(I_{n})+R_{n}$, $T(I_{n})$ = $T(I_{n})+1$
\EndFor
 \State Calculate regret using Eq.~\ref{regret}.
\end{algorithmic}
\end{algorithm}

\begin{algorithm}[!b]
			\caption*{\textbf{Subroutine 3:} ARM\_SEL (Arm Selection Using Chosen Algorithm)}
			\label{amrsel}
			\textbf{Input:}~$alg,n,I_{n-1},X,T,K$\\
			\textbf{Output:}~$I_n$
			\begin{algorithmic}[1]
			\If{$alg==0$} \Comment{UCB Algorithm}
            \State $Q(k,n)= \frac{X(k)}{T(k)} + \sqrt{\frac{\alpha \log(n+K)}{T(k)}}$, $\forall k\in \{1,2,..,K\}$
            \Else \Comment{SBTS-ESSR Algorithm (Subroutine 1)}
            \State Calculate $Q(:,n)$=\textbf{QF\_SBTS\_ESSR}($X$,$T$,$K$,$n$, $I_{n-1}$)
        \EndIf
				\State Select arm, $I_n = \argmax_k Q(:,n)$.
			\end{algorithmic}
		\end{algorithm}

The proposed RI-MAB algorithm is given in Algorithm~\ref{agg_algorithm}.
The number of candidate algorithms, number of arms, and horizon size is $A$, $K$, and $N$, respectively. The RI-MAB algorithm maintains the probability distribution on all candidate algorithms to indicate their optimality. Since all algorithms are equally likely to be optimal at the beginning of each experiment, the prior belief, $\pi_0$, is initialized as uniform distribution (Line 2). To update the belief and identify the optimal algorithm, the RI-MAB algorithm performs epoch-based exploration in the initial $N_{learn}$ time slots (Lines 5-19). In this phase, each algorithm is selected for $2^e$ time slots before incrementing the parameter $e$ by 1 (Line 12-17). This allows a sufficient number of learning samples of each algorithm before finalizing the algorithm for the rest of the horizon i.e. after $N_{learn}$ (Line 20). Such approach allows only one MAB algorithm to be active in hardware in each time slot and increasing length epoch reduces the number of algorithm switching. Even though only one algorithm is active, the parameters, $X$, and $T$, are common across all candidate algorithms which means there is no compromise on arm learning aspects of MAB setup.

During algorithm exploration (Lines 6-18), RI-MAB updates the parameter, $\pi_n$ in each time slot based on the selected algorithm and received reward. Similar to \cite{aggr1}, we obtain the unbiased estimate of received reward using the probability of the arm selection i.e. belief of the selected algorithm (Line 9). Then, the belief of the selected algorithm is updated using the exponential multiplicative factor (Line 10) and learning rate (Line 8)  \cite{aggr1}. In the end, the beliefs of all algorithms are normalized.

The functionality of the RI-MAB algorithm is explained using Fig~\ref{fig:agg_working}. In the beginning, the prior belief, $\pi_0=\{0.5,0.5\}$. The algorithm starts by selecting UCB for the first $2^1=2$ slots. It should be noted that we simulate the algorithm with Bernoulli arm rewards for ease of understanding. In slot 1, UCB selects the arm, $I_1=4$ and receives a reward, $R_1=1$. This results in an increase in the belief value of UCB by a factor of $exp(\eta_{1}*1)$. The same happens in slot 2 when the belief of UCB further increases by receiving $R_2=1$ for the selected arm, $I_2=2$. In slots 3 and 4, the algorithm selects SBTS-ESSR, which receives rewards, $R_3=0$ and $R_4=0$ on arms 4 and 2 respectively. Hence, the belief of TS is not updated in both slots (as $exp(\eta_{n}*0)=1$). This is when the parameter $e$ is incremented by 1. Skipping over to slots 19 and 24, the algorithm selects UCB, which receives a reward, $R_{19}=1$ and $R_{24}=1$ respectively, improving the belief of UCB. In slots 128 and $N_{learn}=500$, the selected candidate algorithm, SBTS-ESSR, receives a reward of 1, which improves its belief. In summary, the belief of the selected candidate algorithm increases when it receives a reward, $R_n=1$, and is not updated when $R_n=0$. After slot $N_{learn}=500$, the algorithm finalizes SBTS-ESSR for the rest of the horizon, pertaining to its higher belief in slot $N_{learn}$.

\begin{figure}[!h]
\includegraphics[width=0.475\textwidth]{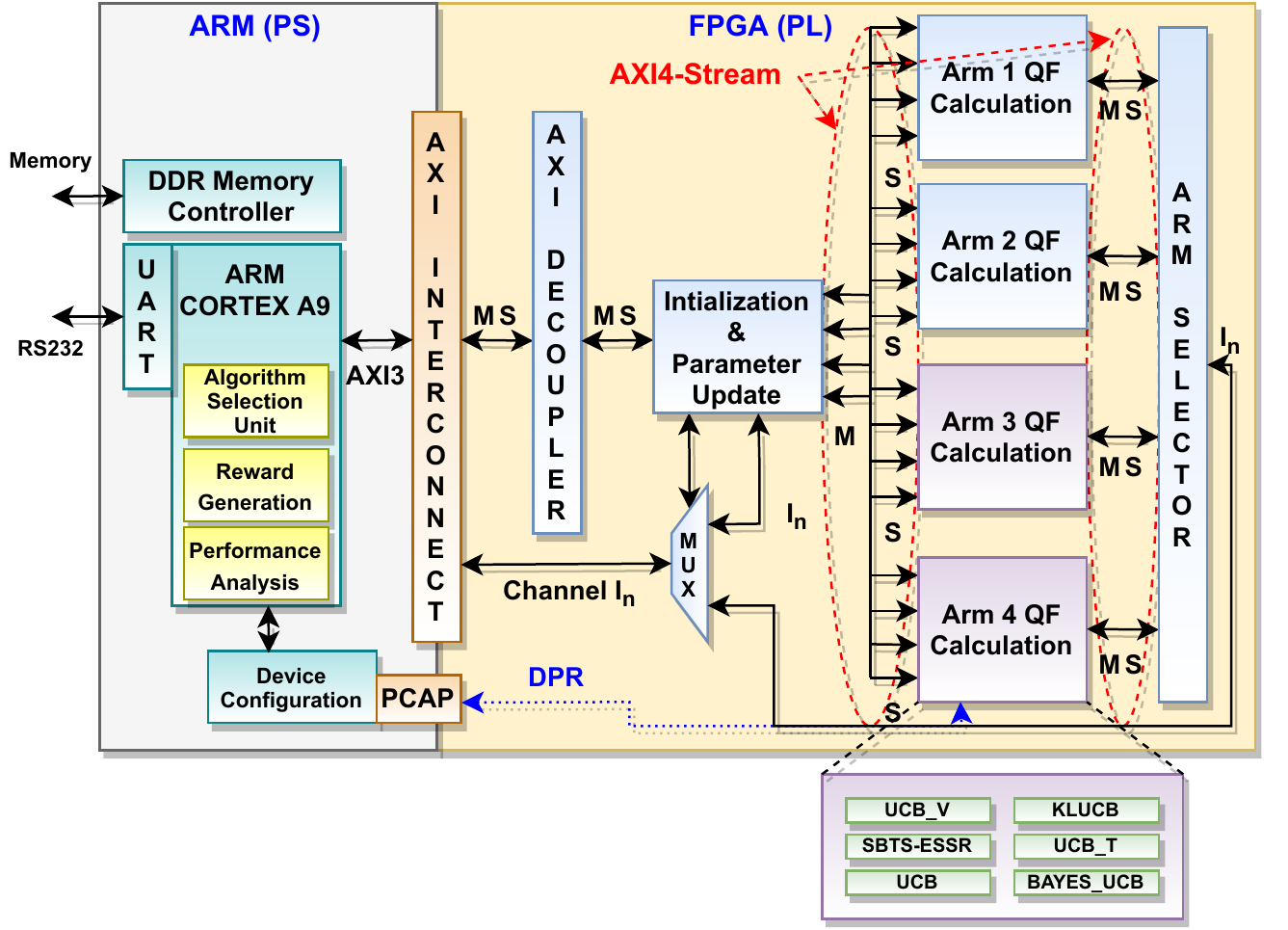}%
\caption{Proposed DPR-enabled architecture for RI-MAB with six candidate algorithms along with reconfigurable $|\beta|$ and $K$.}

\label{main_arch}

\end{figure}

The proposed RI-MAB algorithm is mapped on SoC and the corresponding architecture is shown in Fig.~\ref{main_arch}. Compared to Fig.~\ref{fig:ts_arch}, an additional algorithm selection unit is included in the ARM processor and the QF calculation block is reconfigured via DPR depending on the selected candidate algorithm.

 \begin{figure}[!b]
     \centering
   \includegraphics[width=0.475\textwidth]{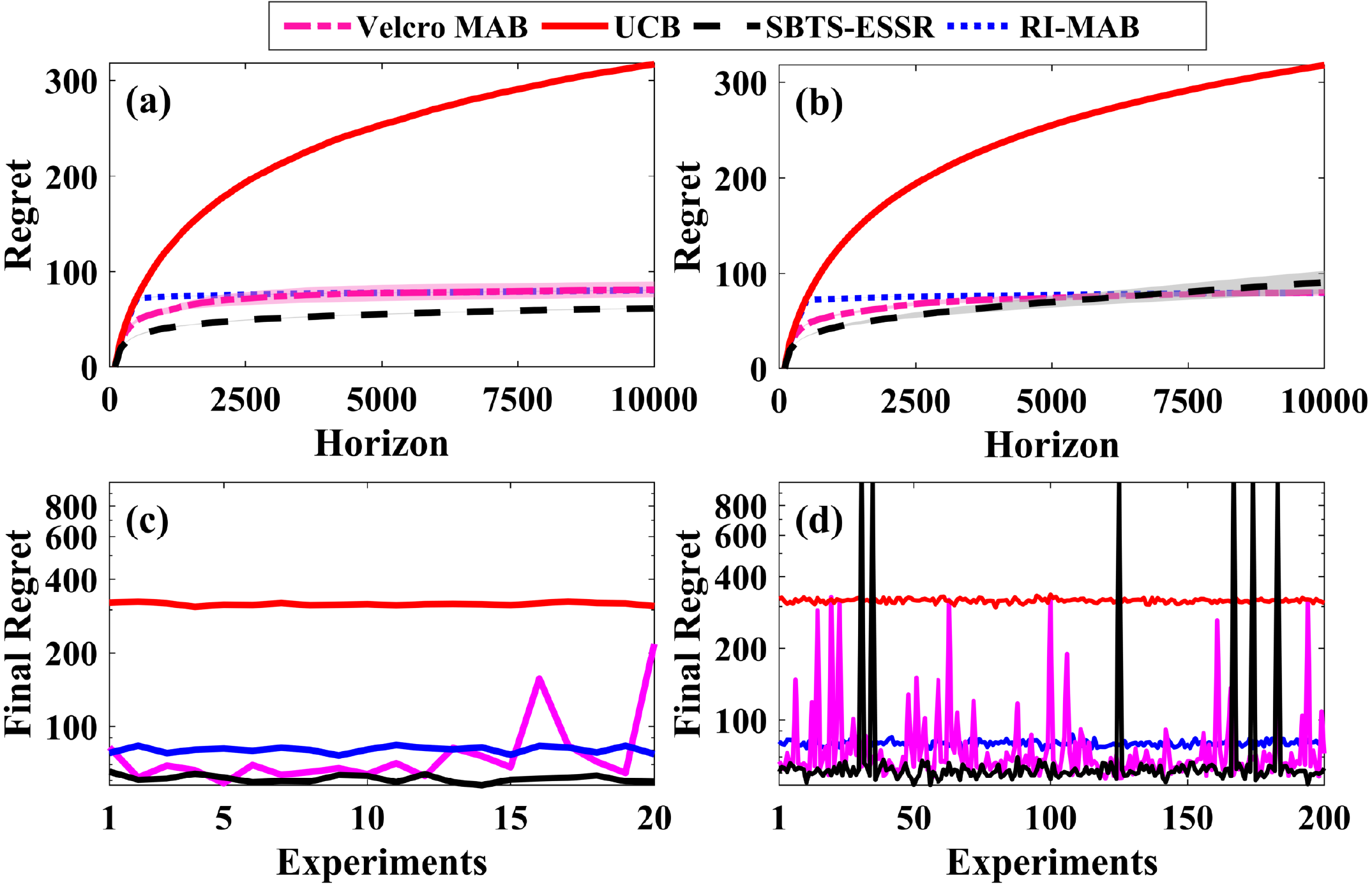}%
     \caption{ Average cumulative regret for $\mu_3$ at the end of (a) 20 experiments, and (b) 200 experiments. Final regret at the end of horizon for (c) 20 experiments, and (d) 200 experiments.}
     \label{fig:aggr_mu3}
 \end{figure}

  \begin{figure*}[!b]
     \centering
   \includegraphics[width=0.9\textwidth]{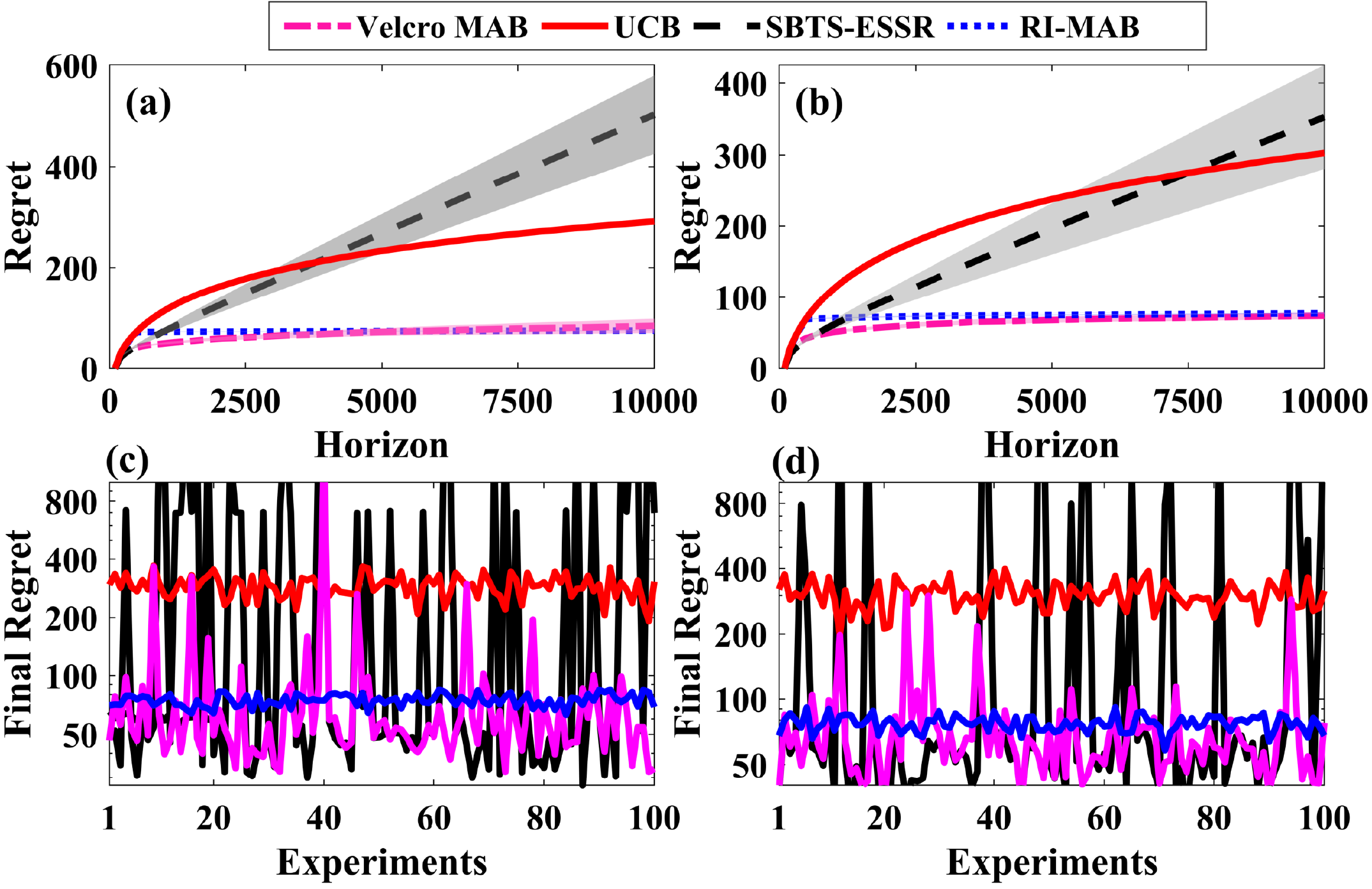}
     \caption{ Average cumulative regret at the end of 100 independent experiments with randomly chosen arm statistics in (a) Case 1, and (b) Case 2. Final regret at the end of horizon for (c) Case 1, and (d) Case 2.}
      \label{fig:aggr_rand}
 \end{figure*}
 
\section{Performance and Complexity Analysis: RI-MAB}
\label{Sec:PRIMAB}
We first begin with the regret comparison of the proposed RI-MAB architecture with the UCB, SBTS-ESSR, and Velcro MAB architectures. To begin with, we consider $N=10000$, $K=8$ and choose $\mu_3$ as the probability distributions. Note that rewards are modeled as Gaussian instead of Bernoulli distribution.

In Fig.~\ref{fig:aggr_mu3} (a), we compare the regret of the four algorithms at different instances of the horizon. The regret is averaged over 20 independent experiments. The final regret (i.e. regret at the end of the horizon) for each experiment is shown in Fig.~\ref{fig:aggr_mu3} (c). In all 20 experiments, SBTS-ESSR offers the lowest regret while UCB incurs the highest regret though both are able to identify the optimal arm. Velcro MAB offers performance closer to SBTS-ESSR in most of the experiments except 16 and 20 where it takes more time to learn TS is better than UCB. On the other hand, RI-MAB selects SBTS-ESSR in all experiments. Despite that, the regret of RI-MAB is higher than Velcro MAB in all experiments except 16 and 30. This happens due to the initial $N_{learn}$ period of exploration to choose between UCB and SBTS-ESSR algorithm. Still, averaged regret of RI-MAB and Velcro-MAB is nearly identical as shown in Fig.~\ref{fig:aggr_mu3} (a) since regret incurred by Velcro-MAB is very high in experiments 16 and 20. Such high regret events are avoided in the RI-MAB algorithm.

In Fig.~\ref{fig:aggr_mu3} (b) and (d), we repeat the above simulations by increasing the number of independent experiments to 200. It can be observed that Velcro MAB and RI-MAB offer lower regret than individual UCB and SBTS-ESSR algorithms. This is because, for experiments 31, 35, 125, 167, 174, 183, SBTS-ESSR fails to identify the optimal arm and hence, incurs huge regret. Also, as expected, UCB incurs slightly higher regret in all experiments. Though Velcro MAB avoids selection of SBTS-ESSR in experiments 31, 35, 125, 167, 174, 183, it wrongly selects UCB in few other experiments (i.e. experiments where the final regret of UCB is same as Velcro MAB). Though RI-MAB incurs higher average regret in experiments where SBTS-ESSR is optimal, it avoids the wrong selection of algorithm as well as optimal arm in the rest of the experiments. It can be observed that Velcro-MAB and RI-MAB offer lower regret than UCB and SBTS-ESSR in some experiments. This happens because learning due to switching between the two algorithms helps to avoid the wrong selection of optimal arm in SBTS-ESSR. These experiments demonstrate the need for careful algorithm switching and selection for a given environment since a single algorithm does not offer optimal performance in all experiments. 
 
\begin{table*}[!b]
\centering
\caption{Comparison of Resource Utilization and Power Consumption}
\label{table:complexity}
\renewcommand{\arraystretch}{1.2}
\resizebox{\textwidth}{!}{
\begin{tabular}{|c|c|c|c|c|c|c|c|c|c|c|c|c|c|c|c|c|}
\hline
\multirow{3}{*}{\textbf{Algorithm}}                                                                                               & \multirow{3}{*}{\textbf{Reconfigurability}}                  & \multicolumn{3}{c|}{\multirow{2}{*}{\textbf{No. of LUTs}}}                                                                                                          & \multicolumn{3}{c|}{\multirow{2}{*}{\textbf{No. of FFs}}}                                                                                                           & \multicolumn{3}{c|}{\multirow{2}{*}{\textbf{No. of DSPs}}}                                                                                                   & \multicolumn{3}{c|}{\multirow{2}{*}{\textbf{No. of BRAMs}}}                                                                                                  & \multicolumn{3}{c|}{\multirow{2}{*}{\textbf{Dynamic Power (W)}}}                                                                                                      \\
                                                                                                                                  &                                                              & \multicolumn{3}{c|}{}                                                                                                                                               & \multicolumn{3}{c|}{}                                                                                                                                               & \multicolumn{3}{c|}{}                                                                                                                                        & \multicolumn{3}{c|}{}                                                                                                                                        & \multicolumn{3}{c|}{}                                                                                                                                                 \\ \cline{3-17} 
                                                                                                                                  &                                                              & \textbf{K=3}                                         & \textbf{K=4}                                         & \textbf{K=5}                                          & \textbf{K=3}                                         & \textbf{K=4}                                         & \textbf{K=5}                                          & \textbf{K=3}                                       & \textbf{K=4}                                       & \textbf{K=5}                                       & \textbf{K=3}                                       & \textbf{K=4}                                       & \textbf{K=5}                                       & \textbf{K=3}                                          & \textbf{K=4}                                          & \textbf{K=5}                                          \\ \hline
\multirow{4}{*}{\textbf{\begin{tabular}[c]{@{}c@{}}RI-MAB,\\ UCB/SBTS-ESSR\\ tunable K $\leq$ $K_{max}$\\ $|\beta|=20$\end{tabular}}} & \textbf{\begin{tabular}[c]{@{}c@{}}DPR\\ SP-FP\end{tabular}} & \begin{tabular}[c]{@{}c@{}}7664\\ \textbf{-65\%}\end{tabular} & \begin{tabular}[c]{@{}c@{}}9866\\ \textbf{-55\%}\end{tabular} & \begin{tabular}[c]{@{}c@{}}12068\\ \textbf{-44\%}\end{tabular} & \begin{tabular}[c]{@{}c@{}}7060\\ \textbf{-59\%}\end{tabular} & \begin{tabular}[c]{@{}c@{}}9093\\ \textbf{-48\%}\end{tabular} & \begin{tabular}[c]{@{}c@{}}11126\\ \textbf{-36\%}\end{tabular} & \begin{tabular}[c]{@{}c@{}}51\\ \textbf{-56\%}\end{tabular} & \begin{tabular}[c]{@{}c@{}}68\\ \textbf{-41\%}\end{tabular} & \begin{tabular}[c]{@{}c@{}}85\\ \textbf{-27\%}\end{tabular} & \begin{tabular}[c]{@{}c@{}}27\\ \textbf{-49\%}\end{tabular} & \begin{tabular}[c]{@{}c@{}}36\\ \textbf{-32\%}\end{tabular} & \begin{tabular}[c]{@{}c@{}}45\\ \textbf{-15\%}\end{tabular} & \begin{tabular}[c]{@{}c@{}}0.127\\ \textbf{-65\%}\end{tabular} & \begin{tabular}[c]{@{}c@{}}0.165\\ \textbf{-54\%}\end{tabular} & \begin{tabular}[c]{@{}c@{}}0.203\\ \textbf{-44\%}\end{tabular} \\ \cline{2-17} 
                                                                                                                                  & \textbf{\begin{tabular}[c]{@{}c@{}}DPR\\ WL=27\end{tabular}} & \begin{tabular}[c]{@{}c@{}}5761\\ \textbf{-74\%}\end{tabular} & \begin{tabular}[c]{@{}c@{}}7375\\ \textbf{-66\%}\end{tabular} & \begin{tabular}[c]{@{}c@{}}8989\\ \textbf{-59\%}\end{tabular}  & \begin{tabular}[c]{@{}c@{}}4390\\ \textbf{-75\%}\end{tabular} & \begin{tabular}[c]{@{}c@{}}5574\\ \textbf{-68\%}\end{tabular} & \begin{tabular}[c]{@{}c@{}}6758\\ \textbf{-61\%}\end{tabular}  & \begin{tabular}[c]{@{}c@{}}0\\ \textbf{-100\%}\end{tabular} & \begin{tabular}[c]{@{}c@{}}0\\ \textbf{-100\%}\end{tabular} & \begin{tabular}[c]{@{}c@{}}0\\ \textbf{-100\%}\end{tabular} & \begin{tabular}[c]{@{}c@{}}3\\ \textbf{-95\%}\end{tabular}  & \begin{tabular}[c]{@{}c@{}}4\\ \textbf{-93\%}\end{tabular}  & \begin{tabular}[c]{@{}c@{}}5\\ \textbf{-91\%}\end{tabular}  & \begin{tabular}[c]{@{}c@{}}0.019\\ \textbf{-95\%}\end{tabular} & \begin{tabular}[c]{@{}c@{}}0.023\\ \textbf{-94\%}\end{tabular} & \begin{tabular}[c]{@{}c@{}}0.027\\ \textbf{-93\%}\end{tabular} \\ \cline{2-17} 
                                                                                                                                  & \textbf{\begin{tabular}[c]{@{}c@{}}DPR\\ WL=11\end{tabular}} & \begin{tabular}[c]{@{}c@{}}2425\\ \textbf{-89\%}\end{tabular} & \begin{tabular}[c]{@{}c@{}}2927\\ \textbf{-87\%}\end{tabular} & \begin{tabular}[c]{@{}c@{}}3429\\ \textbf{-84\%}\end{tabular}  & \begin{tabular}[c]{@{}c@{}}2443\\ \textbf{-86\%}\end{tabular} & \begin{tabular}[c]{@{}c@{}}2978\\ \textbf{-83\%}\end{tabular} & \begin{tabular}[c]{@{}c@{}}3513\\ \textbf{-80\%}\end{tabular}  & \begin{tabular}[c]{@{}c@{}}0\\ \textbf{-100\%}\end{tabular} & \begin{tabular}[c]{@{}c@{}}0\\ \textbf{-100\%}\end{tabular} & \begin{tabular}[c]{@{}c@{}}0\\ \textbf{-100\%}\end{tabular} & \begin{tabular}[c]{@{}c@{}}3\\ \textbf{-95\%}\end{tabular}  & \begin{tabular}[c]{@{}c@{}}4\\ \textbf{-93\%}\end{tabular}  & \begin{tabular}[c]{@{}c@{}}5\\ \textbf{-91\%}\end{tabular}  & \begin{tabular}[c]{@{}c@{}}0.013\\ \textbf{-97\%}\end{tabular} & \begin{tabular}[c]{@{}c@{}}0.015\\ \textbf{-96\%}\end{tabular} & \begin{tabular}[c]{@{}c@{}}0.017\\ \textbf{-95\%}\end{tabular} \\ \cline{2-17} 
                                                                                                                                  & \textbf{Velcro (SP-FP)}                                      & \multicolumn{3}{c|}{21563}                                                                                                                                          & \multicolumn{3}{c|}{17201}                                                                                                                                          & \multicolumn{3}{c|}{115}                                                                                                                                     & \multicolumn{3}{c|}{52.5}                                                                                                                                    & \multicolumn{3}{c|}{0.358}                                                                                                                                            \\ \hline
\end{tabular}
}
\end{table*}

Next, we select the arm statistics randomly in each of the experiments instead of fixed $\mu_3$ in Fig.~\ref{fig:aggr_mu3}. We consider $N=10000$, $K=8$ and two types of arm statistics such that the minimum difference between two arm statistics is 1) 0.07 (Case 1), and 2) 0.025 (Case 2). Corresponding results are shown in Fig.~\ref{fig:aggr_rand} (a)-(d). It can be observed that SBTS-ESSR offers significantly higher regret in almost 25\% of the experiments. Though UCB offers an overall higher average regret, it successfully identifies optimal arm in all experiments leading to better performance than SBTS-ESSR. Proposed RI-MAB offers similar performance as UCB but with lower average regret. On the other hand, Velcro UCB fails to identify the appropriate algorithm and hence, the optimal arm in some experiments while in other experiments it offers lower regret than RI-MAB. On average, both offer nearly identical average regret but RI-MAB eliminates high regret events.

\begin{figure}[!h]
     \centering
   \includegraphics[width=0.5\textwidth]{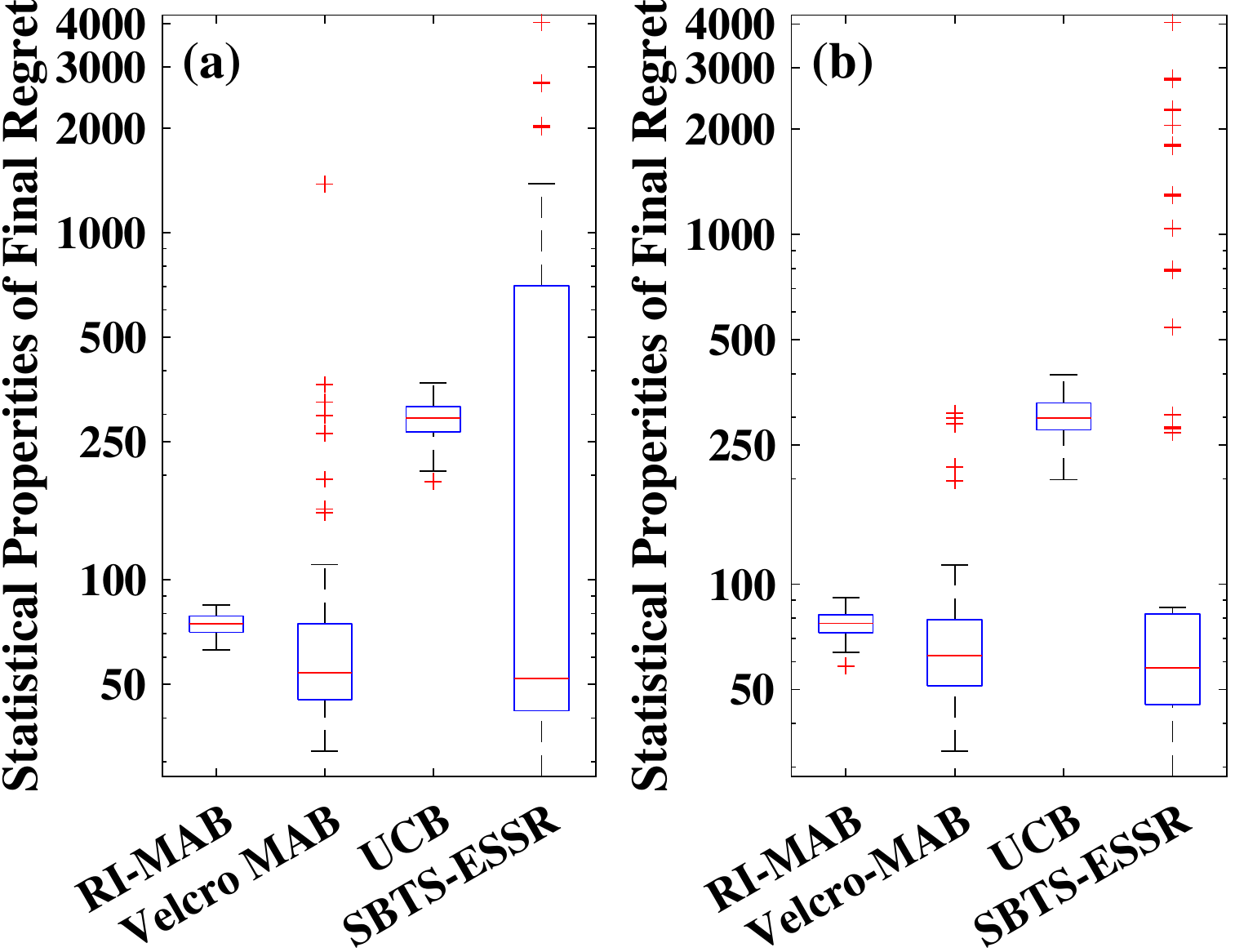}%
     \caption{ Statistical properties of the final regret of the different algorithms at the end of 100 independent experiments with randomly chosen arm statistics in (a) Case 1, and (b) Case 2.}
      \label{fig:aggr_stats}
 \end{figure}

To gain further insights into the regret performance, we study various statistical properties of the final regret over 100 experiments in Fig.~\ref{fig:aggr_stats}. Using the Boxplot feature in Matlab, we consider median (central red line), and percentile (bottom and top edges of the box indicate the 25th and 75th percentiles, respectively). The outliers are plotted individually using the '+' symbol. It can be observed that the SBTS-ESSR and Velco-MAB have large size boxes indicating large variation in regret performance and the number of outliers are more indicating wrong selection of algorithm or arm. On the other hand, UCB and RI-MAB guarantee the selection of optimal arm in all experiments and RI-MAB offers lower regret between them.

Next, we compare the resource utilization and power consumption of RI-MAB and Velcro-MAB architectures in Table~\ref{table:complexity}. Both architectures are on-the-fly reconfigurable in terms of the number of arms and type of the algorithm. Proposed RI-MAB with SP-FP WL architecture offers around 44-65\%, 36-59\%, 27-56\% and 15-49\% savings in LUTs, FFs, DSPs, and BRAMs over Velcro-MAB architecture. Similarly, it offers 44-65\% lower dynamic power consumption.  The savings improve further when the WL is optimized via fixed-point representation. With the increase in the number of arms (i.e. for $K_{max} > 20$, the proposed architecture and DPR approach further improvements in resource utilization. In addition, savings will increase further with the increase in the number of candidate algorithms.

\section{Conclusions and Future Works}
\label{Sec:Conclusion}
In this paper, we present synthesizable and reconfigurable architecture of the Thompson Sampling (TS) multi-armed bandit (MAB) algorithm for arms with Bernoulli distribution. The functional correctness, execution time, resource, and power consumption comparisons demonstrate its superiority over the upper confidence bound (UCB) algorithm. For an environment with unknown arm distribution, a reconfigurable and intelligent MAB (RI-MAB) algorithm is proposed along with its architecture. The RI-MAB offers significant savings in resources and power consumption without compromising on the regret performance. In the future, we plan to integrate the proposed RI-MAB with wireless radio and analyze the gain in throughput using real radio signals. Other possibilities include the extension of the RI-MAB architecture for an environment where the number of arms, as well as arm statistics, are not fixed i.e. the non-stationary environment in which an optimal arm changes over time.

\bibliographystyle{IEEEtran}
\bibliography{main.bbl}

\end{document}